\documentclass[nofootinbib,amsmath,amssymb,showpacs,showkeys,aps,reprint]{revtex4-1}
\usepackage[utf8,latin1]{inputenc}
\usepackage{graphicx}
\usepackage{dcolumn}
\usepackage[dvipsnames]{xcolor}
\usepackage[T1]{fontenc}

\usepackage{mathrsfs}  
\usepackage{cases}
\usepackage{bm}
\usepackage{academicons}
\usepackage{mathtools, nccmath}
\usepackage{fancyhdr}
\usepackage{tikz,xcolor}
\newcommand{\qm}[1]{``#1''}
\usepackage{tensor}
\usepackage[normalem]{ulem}
\usepackage{lipsum}
\usepackage{soul}
\usepackage{cancel}
\usepackage{stackengine,scalerel}
\usepackage{mathabx}
\usepackage{hyperref}
\usepackage{tabularx}
\hypersetup{colorlinks, linkcolor={blue},citecolor={blue},urlcolor={blue}}

\usepackage{comment}
\usepackage{accents}
\usepackage{cleveref}
\usepackage{physics}
\usepackage{amssymb}

\newcommand\overstarbf[1]{\ThisStyle{\ensurestackMath{%
  \stackengine{0pt}{\SavedStyle\mathbf{#1}}{\smash{\SavedStyle*}}{O}{c}{F}{T}{S}}}}

\definecolor{lime}{HTML}{A6CE39}
\DeclareRobustCommand{\orcidicon}{
	\begin{tikzpicture}
	\draw[lime, fill=lime] (0,0) 
	circle [radius=0.16] 
	node[white] {{\fontfamily{qag}\selectfont \tiny ID}};
	\draw[white, fill=white] (-0.0625,0.095) 
	circle [radius=0.007];
	\end{tikzpicture}
	\hspace{-2mm}
}

\def\nn{\nonumber}

\fancypagestyle{plain}{%
  \fancyhf{}
  \fancyfoot[C]{\iffloatpage{}{\thepage}}
  }
\foreach \x in {A, ..., Z}{%
	\expandafter\xdef\csname orcid\x\endcsname{\noexpand\href{https://orcid.org/\csname orcidauthor\x\endcsname}{\noexpand\orcidicon}}
}

\newcommand\restr[2]{{% we make the whole thing an ordinary symbol
  \left.\kern-\nulldelimiterspace % automatically resize the bar with \right
  #1 % the function
  \littletaller % pretend it's a little taller at normal size
  \right|_{#2} % this is the delimiter
  }}

\newcommand{\littletaller}{\mathchoice{\vphantom{\big|}}{}{}{}}

\begin{document}

\title[Rotating Weyssenhoff fluids in linearized Einstein-Cartan theory]{Rotating Weyssenhoff fluids in linearized Einstein-Cartan theory}

\author{Emmanuele Battista\orcidC{}$^{1}$\vspace{0.5cm}}\email{ebattista@lnf.infn.it}%\email{emmanuelebattista@gmail.com}
\author{Salvatore Capozziello\orcidA{}$^{2,3,4}$}
\email{capozziello@na.infn.it}
\author{Francesco Giovinetti\orcidB{}$^{2,3}$} \email{francesco.giovinetti@unina.it}

\affiliation{$^1$ Istituto Nazionale di Fisica Nucleare, Laboratori Nazionali di Frascati, 00044 Frascati, Italy\\
$^2$ Dipartimento di Fisica ``Ettore Pancini'', Complesso Universitario 
di Monte S. Angelo, Universit\`a degli Studi di Napoli ``Federico II'', Via Cintia Edificio 6, 80126 Napoli, Italy\\
$^3$ Istituto Nazionale di Fisica Nucleare, Sezione di Napoli, Complesso Universitario 
di Monte S. Angelo, Via Cintia Edificio 6, 80126 Napoli, Italy
\\
$^4$ Scuola Superiore Meridionale, Largo San Marcellino 10, 80138 Napoli, Italy 
}

%\date{\today} 

\begin{abstract}

The Weyssenhoff fluid expands the ordinary perfect-fluid description of general relativity by taking into account the  intrinsic spin of its constituent particles. Within the linearized regime of Einstein-Cartan theory,  we investigate the exterior geometry produced by isolated slowly rotating Weyssenhoff fluids, i.e.,  sources endowed with both macroscopic and microscopic angular momenta. Our work represents an extension of the foundational analysis of Arkuszewski, Kopczy{\'n}ski and Ponomariev, which was restricted to the static fluid configuration. Under the assumptions of stationarity and axial symmetry, we derive the explicit expression for the linearized stress-energy tensor of the source and obtain the corresponding metric  through a multipole expansion. The ensuing solution exhibits a generalized Lense-Thirring structure, with both the gravitoelectric and gravitomagnetic sectors affected by the quantum spin and macroscopic rotation. We also discuss some phenomenological implications of the model,  including the gyroscope precession  and the Sagnac effect, along with potential experimental probes of the predicted corrections.

\end{abstract}

\maketitle

%\tableofcontents

\section{Introduction}

General relativity (GR) has achieved remarkable success over a broad range of domains, ranging from 
Solar-System observations \cite{Will2014} to the direct detection of gravitational waves \cite{Blanchet2013}, the imaging of black hole shadows \cite{EventHorizonTelescope2019}, and the standard cosmological model \cite{Peebles2002}.  Nevertheless, several theoretical considerations suggest that the underlying Riemannian geometry is sufficient to provide a correct description of gravity only in the macroscopic regime, while difficulties arise when moving to the microscopic scales.  Among the various  motivations for going beyond the GR paradigm, the problem of quantizing the gravitational field is perhaps the most compelling. Simple power counting arguments indicate that GR is  perturbatively non-renormalizable, since the Newtonian constant $G$ has  negative mass dimension (in natural units $\hbar=c=1$), leading to  ultraviolet divergences beyond one-loop level  that cannot be canceled  via a redefinition of the field parameters   \cite{tHooft1974,Goroff1985,Hamber2009}. Furthermore, the well-known Hawking-Penrose theorems prove that GR is plagued by spacetime singularities, which suggest that  the universe originated from a singular state and that  the final outcome of gravitational collapse may be pathological  \cite{Hawking-Penrose1970,Senovilla1998}.

Abandoning Riemannian geometry  provides a first attempt to address  these difficulties.  
The simplest extension of GR consists  in passing to the Riemann-Cartan spacetime, where the affine connection  can acquire  torsion contributions \cite{Hehl:1976kj,Medina2018}. This  framework gives rise to  Einstein-Cartan (EC) gravity, which can be formulated as a gauge theory of the Poincar\'e group and thus represents a particular case of the broader setting of metric-affine (gauge theory of) gravity \cite{Hehl1994-MAG,Cano2021}. EC theory allows to overcome a conceptual limitation of GR concerning  microstructured matter. According to quantum field theory,  particles can be classified by the irreducible unitary representations of the Poincar\'e group and are characterized by two fundamental quantities, namely the mass and spin. While mass(-energy) acts as the source of the spacetime curvature in GR,  spin  does not fulfill an independent  geometrical role.  This shortcoming is resolved in EC gravity, where torsion is naturally coupled (in an algebraic way) to the spin of the matter. Here, we adhere to the standard EC interpretation, and by \textit{spin} we mean the \textit{quantum intrinsic angular momentum}  carried by elementary particles \cite{Hehl:1976kj,Bohmer2017}, as opposed   to the classical orbital angular momentum associated with the macroscopic rotation of a body.

EC theory  finds applications in a variety of physical scenarios.   From a field-theoretic standpoint,   torsion can provide a regularization prescription for ultraviolet-divergent integrals in loop Feynman diagrams \cite{Poplawski:2017yeo}, and  induce nontrivial interactions with  matter fields, particularly in its coupling to fermions \cite{Poplawski2009,Cabral:2019gzh,Karananas2021} (see also Refs. \cite{Battista:2022vvl,Capolupo:2023igw}). From a foundational perspective, singularity theorems and focusing conditions have  recently  been examined in detail \cite{Luz2018,Luz2019,Luz2023,Choudhury2024}, partly motivated by the possibility that singularities  may be avoided under suitable circumstances in gravitational collapse and black hole formation settings \cite{Ziaie:2013pma,Hashemi2014,Poplawski:2020pie}, as well as in cosmology \cite{deBerredo-Peixoto2009,Poplawski:2013koa}. In the latter context, torsion can generate nonsingular bouncing models \cite{Poplawski:2012ab,Cubero2019,Poplawski:2020hrp}, affect inflationary dynamics \cite{Poplawski2010,Poplawski:2018ypb,Piani2022,Piani2023}, account for the  matter-antimatter asymmetry problem \cite{Poplawski:2011xf},  and  influence the late-time evolution of the universe,  leading to departures from standard dark sector phenomenology \cite{Bolejko2020}. On the astrophysical side, EC gravity has  been applied to the study of gravitational waves \cite{Elizalde2022,Battista:2021rlh,Battista:2022hmv,DeFalco:2024ojf,Barriga2025},    post-Newtonian dynamics of compact binaries \cite{Battista:2022sci,Battista:2023znv,DeFalco:2023djo}, and neutron star interiors \cite{Jockel2024}. 

A number of investigations, including many of those mentioned above, rely on the \textit{Weyssenhoff fluid}, which generalizes the ordinary concept of perfect fluid and provides an effective macroscopic description of matter possessing intrinsic spin \cite{Trautman1973,Hehl:1976kj,Obukhov:1987yu,Boehmer:2006gd,Bohmer2017}.
In this framework, a suitable averaging procedure over the microscopic spin degrees of freedom is adopted and  a collection of  constituents with a random spin distribution is treated as a continuous medium. As a result, the average spin density tensor vanishes, while quadratic-in-spin terms survive and act as a source of torsion in the EC field equations \cite{deBerredo-Peixoto2009}.

The main deviations from GR are expected to arise at extremely high matter densities, with critical values of order $10^{50}$--$10^{57}\,\mathrm{kg/m^3}$ \cite{Hehl:1976kj}. These severe conditions, which  
are typically relevant  in the early universe and final stages of gravitational  collapse,  make the detection of torsion effects challenging. Despite that, various   empirical tests of EC theory have been proposed in the literature  (see, e.g., the overview of possible measurement schemes and their limitations in Ref. \cite{Hehl:2013qga}). These include, in particular,  spin-precession and polarization experiments probing torsion-induced spin couplings \cite{Obukhov:2014fta}, which can be formulated even  within the weak-field approximation \cite{ArderucioCosta:2023ooz}, potential modifications of electromagnetic wave propagation \cite{Bolejko2020,Trukhanova:2022utt},  as well as high-energy signatures based on effective fermion-torsion interactions, see e.g. Ref.  \cite{Shapiro:2001rz} for a review.

Motivated by the importance of  identifying potential scenarios where torsion might play a significant role and of devising possible experimental setups to evaluate its existence, in this paper we analyze  rotating  Weyssenhoff fluids, i.e., sources endowed with \emph{both} intrinsic quantum spin \emph{and} orbital angular momentum, within the linearized EC approximation. This framework, to the best of our knowledge, is presented here for the first time. 

The study of the exterior spacetime of a Weyssenhoff fluid sphere in linearized EC gravity was initiated by  Arkuszewski, Kopczy{\'n}ski and Ponomariev in Ref. \cite{arkuszewski1974linearized}, which deals with the \textit{static} configuration where
all fluid elements are at rest with respect to an asymptotically inertial  frame, i.e.,  the source has no macroscopic rotation. The  external metric is found to be formally analogous to the Lense-Thirring solution \cite{MTW1973}, with the off-diagonal components determined by the total intrinsic spin of the Weyssenhoff fluid, rather than by the classical angular momentum. 

As we will show, when  macroscopic rotation is switched on,   the metric attains a generalized Lense-Thirring form at  the lowest nontrivial multipole order. Both the gravitoelectric and gravitomagnetic sectors exhibit peculiar corrections: the former acquires an additional \qm{spin-orbit} contribution whose sign is controlled by the relative orientation of the intrinsic spin and orbital angular momentum of the fluid, while the latter depends on the linear combination of the classical and quantum spins. 

The plan of the paper is thus as follows. After reviewing the main aspects of EC gravity and its linearized limit in Sec. \ref{sec:Linearized Einstein-Cartan theory}, we  examine  the external weak-field geometry of an  isolated, stationary, axisymmetric, and slowly rotating  Weyssenhoff fluid  in Sec. \ref{sec:Weyssenhoff_fluid_source_and_solution}. Our approach proceeds by first computing the (metric) stress-energy tensor of the source in accordance with its symmetries, and then  determining the external metric by performing a multipole expansion of the Poisson-integral solution of the linearized EC field equations, followed by an analysis of the junction conditions at the fluid-vacuum interface. In Sec. \ref{sec:tests-GEM}, we discuss the phenomenological aspects of the solution and potential tests by studying  gravitomagnetic observables like the Lense-Thirring precession and Sagnac effect, as well as some spin-rotation coupling mechanisms  such as the Barnett and Einstein-de Haas phenomena. Finally,  in Sec. \ref{Sec:Conclusion} we draw our conclusions.

\emph{Notations and conventions.} We use metric signature  $(-,+,+,+)$. Greek indices take values  $0,1,2,3$, while lowercase Latin ones $1,2,3$.  Four-vectors are written as $a^\mu = (a^0,a^i)$.  Spatial three-vectors are denoted by $\vec{a}=(a^1,a^2,a^3)$, with $\vec{a}\cdot\vec{b}:=\delta_{lk}a^l b^k$, $\vert \vec{a}\vert:=(\vec{a}\cdot\vec{a})^{1/2} \equiv a$, and $(\vec{a}\times\vec{b})^i:=\epsilon_{ilk}a^l b^k$. An overhat symbol refers to quantities defined in terms of the Levi-Civita connection.  The symmetrization and antisymmetrization procedures read as $A_{(\mu \nu)}=\frac{1}{2}(A_{\mu \nu}+A_{\nu \mu})$ and $A_{[\mu \nu]}=\frac{1}{2}(A_{\mu \nu}-A_{\nu \mu})$, respectively.

\section{Einstein-Cartan gravity and its weak-field limit}\label{sec:Linearized Einstein-Cartan theory}

The EC model represents the simplest generalization of GR that  allows for a torsionful connection. Since the underlying action is formally identical to the Einstein-Hilbert action of GR,   torsion has no dynamical character  and vanishes in vacuum, where the theory boils down  to GR \cite{Hehl:1976kj,Medina2018} (for further details, we refer the reader to Refs. \cite{Gasperini-DeSabbata1985,Capozziello:2001mq,Trautman:2006fp,Poplawski:2009fb,Blagojevic2013,Ponomarev2017,Diether2017,ArderucioCosta:2023ooz,Luz2025,Pastore2026}). 

In this section, we briefly outline the main aspects of EC gravity relevant for our analysis. After introducing the  geometric framework and setting our conventions in Sec. \ref{Sec:mathematical-preliminaries}, we discuss the EC field equations in Sec. \ref{Sec:EC-theory}. Then, the  linearized  regime is presented in Sec. \ref{Sec:linearized-EC-theory}.

\subsection{The geometric framework}\label{Sec:mathematical-preliminaries}

EC gravity   entails the introduction of a four-dimensional Riemann-Cartan spacetime manifold $\mathcal{M}$ endowed with a Lorentzian-signature metric  $g_{\mu \nu}$  and a metric-compatible affine connection $\tensor{\Gamma}{^\lambda_{\mu\nu}}$, which admits the decomposition
\begin{align}
\tensor{\Gamma}{^\lambda_{\mu\nu}}=\tensor{\hat{\Gamma}}{^\lambda_{\mu\nu}}-\tensor{K}{_{\mu\nu}^\lambda}. 
\end{align}
Here,   $\tensor{\hat{\Gamma}}{^\lambda_{\mu\nu}}$ denotes the Levi-Civita connection 
\begin{align}
\tensor{\hat{\Gamma}}{^\lambda_{\mu\nu}}:=
\frac{1}{2}g^{\lambda\alpha}\left( \partial_{\mu}g_{\nu\alpha} + \partial_{\nu}g_{\alpha \mu} -\partial_{\alpha}g_{\mu\nu}\right),\label{eq:def christoffel}
\end{align}
while 
\begin{align}
\tensor{K}{_{\mu\nu}^\lambda}\coloneq&-\tensor{S}{_{\mu\nu}^\lambda}+\tensor{S}{_\nu^\lambda_\mu}-\tensor{S}{^\lambda_{\mu\nu}}=-\tensor{K}{_{\mu}^\lambda_{\nu}},\label{eq:controsion definition}    
\end{align}
represents the \textit{contorsion tensor}, with 
\begin{align}
    \tensor{S}{_{\mu\nu}^{\lambda}}:=\tensor{\Gamma}{^\lambda_{\left[\mu\nu\right]}}=-\tensor{S}{_{\nu\mu}^{\lambda}},
\end{align}
the \textit{torsion tensor}.

In our conventions,  the Riemann tensor reads as \cite{Hehl:1976kj} 
\begin{align}
\tensor{R}{_{\mu\nu\lambda}^{\rho}}=&\partial_\mu\tensor{\Gamma}{^\rho_{\nu\lambda}}-\partial_\nu\tensor{\Gamma}{^\rho_{\mu\lambda}}+\tensor{\Gamma}{^\rho_{\mu\eta}}\tensor{\Gamma}{^\eta_{\nu\lambda}}-\tensor{\Gamma}{^\rho_{\nu\eta}}\tensor{\Gamma}{^\eta_{\mu\lambda}},
\label{eq:def Riemann tensor} 
\end{align}
and the Einstein tensor assumes    the usual form
\begin{align}
    \tensor{G}{_{\mu\nu}}\coloneq \tensor{R}{_{\mu\nu}}-\frac{1}{2}g_{\mu\nu}R, 
\end{align}
the Ricci tensor and Ricci  scalar being given by $R_{\mu\nu}:=\tensor{R}{_{\alpha\nu\mu}^\alpha}$  and $R:=g^{\mu\nu}R_{\mu\nu}$, respectively.

\subsection{The field equations}\label{Sec:EC-theory}

Since in the EC framework the gravitational field has two geometric
degrees of freedom, namely curvature and torsion, its description requires the presence of two sets of field equations.  These can be derived from the  total action  
\begin{align}
\mathscr{S}=\frac{1}{c}\int\mathrm{d}^4x\sqrt{- g}\left(\frac{1}{2\chi}R+\mathcal{L}_m \right),\label{eq:total action EC}
\end{align}
where $\chi:= 8\pi G/c^4$  and the matter Lagrangian $\mathcal{L}_m$ is assumed to be minimally coupled to gravity (see, e.g., Refs. \cite{Cembranos2017,Chen:2018szr,Ovgun:2020gjz,Pantig2022b}  for extensions involving generalized gravitational actions). Varying  $\mathscr{S}$ with respect to   $g_{\mu \nu}$ and $\tensor{K}{_{\mu\nu}^\lambda}$ yields the EC field equations (with the one for the metric  being symmetric, as expected from the symmetry of $g_{\mu\nu}$). Using the (generalized) Belinfante-Rosenfeld relation, the resulting system  can be equivalently recast in the  form \cite{Gasperini-DeSabbata1985}
\begin{subequations}
\label{EC-equations-1}
\begin{align}
\tensor{G}{^{\alpha\beta}}=&\chi\tensor{\mathbb{T}}{^{\alpha\beta}}
\label{eq:field eqs energy momentum R},\\
\tensor{S}{_{\alpha\beta}^\gamma}+2\tensor{\delta}{^\gamma_{\left[\alpha\right.}}\tensor{S}{_{\beta\left.\right]\lambda}^\lambda}=&\chi\tensor{\tau}{_{\alpha\beta}^\gamma}.
\label{eq:field eqs spin R}
\end{align}
\end{subequations}
Here,  $\tensor{\mathbb{T}}{^{\alpha\beta}}$ is the (asymmetric) \textit{canonical energy-momentum tensor},  which is related to the (symmetric) \textit{metric energy-momentum tensor}
\begin{align}
\sqrt{-g}\,\tensor{T}{^{\alpha\beta}}&=2\frac{\delta\left(\sqrt{-g}\mathcal{L}_m\right)}{\delta g_{\alpha\beta}},\label{eq:def dyn canon energy momentum}
\end{align}
via the aforementioned (generalized) Belinfante-Rosenfeld identity
\begin{align}
\tensor{\mathbb{T}}{^{\alpha\beta}}&= T^{\alpha\beta}-\overstarbf{\nabla}_\gamma\left( -\tau^{\alpha\beta\gamma}+\tau^{\beta\gamma\alpha}-\tau^{\gamma\alpha\beta}\right), \label{bold-T-tensor}
\end{align}
with $\tensor{\accentset{*}{\nabla}}{_\alpha}:=\tensor{\nabla}{_\alpha}+2\tensor{S}{_{\alpha\beta}^\beta}$, while  
\begin{align}
\sqrt{-g}\,\tensor{\tau}{_{\alpha\beta}^\gamma}&=\frac{\delta\left(\sqrt{-g}\mathcal{L}_m\right)}{\delta   \tensor{K}{_\gamma^\beta^\alpha}},\label{eq: def dyn tau 2}
\end{align}
denotes the \textit{spin angular momentum tensor}.

Starting from  Eq. \eqref{EC-equations-1}, one can derive the relations \cite{Hehl:1976kj,Gasperini-DeSabbata1985}
\begin{align}
&\tensor{\accentset{*}{\nabla}}{_\nu}\tensor{\mathbb{T}}{_\mu^\nu}=2\tensor{\mathbb{T}}{_\lambda^\nu}\tensor{S}{_{\mu\nu}^\lambda}+\tensor{\tau}{_{\nu\rho}^\sigma}\tensor{R}{_{\mu\sigma}^{\nu\rho}},\label{eq:conserv energy momentum}\\
&\tensor{\accentset{*}{\nabla}}{_\gamma}\tensor{\tau}{_{\alpha\beta}^\gamma}-\tensor{\mathbb{T}}{_{\left[\alpha\beta\right]}}=0,\label{eq:conserv angular momentum}
\end{align}
which represent the generalized version of the energy-momentum and  total (spin plus orbital) angular momentum conservation laws, respectively. 

It follows from the above formulas that the mass-energy and spin densities of matter, which act as sources of the gravitational field, are geometrically related to the spacetime  curvature and torsion, respectively.  Another   key aspect of EC gravity is that Eq. \eqref{eq:field eqs spin R} is an algebraic rather than a dynamical equation. This  feature has a twofold consequence. First, the torsion  can be  formally eliminated in favor of the spin tensor in the field equations, which can be thus   rewritten in the GR-like form 
\begin{equation}\label{eq:field eqs for metric G}
\tensor{\hat{G}}{^{\mu\nu}}=\chi \tensor{T}{^{\mu\nu}}+\chi^2\tensor{\tilde{\tau}}{^{\mu\nu}},
\end{equation}
 $\tilde{\tau}^{\mu \nu}$ being the symmetric tensor
\begin{align}
{\tilde{\tau}}^{\mu\nu}:=&-4\tensor{\tau}{^{\mu\gamma}_{[\lambda}}\tensor{\tau}{^{\nu\lambda}_{\gamma]}}-2\tensor{\tau}{^{\mu\gamma\lambda}}\tensor{\tau}{^{\nu}_{\gamma\lambda}}+\tensor{\tau}{^{\gamma\lambda\mu}}\tensor{\tau}{_{\gamma\lambda}^{\nu}}\nonumber\\
&+\frac{1}{2}g^{\mu\nu}\left(4\tensor{\tau}{_{\alpha}^{\gamma}_{[\lambda}}\tensor{\tau}{^{\alpha\lambda}_{\gamma]}}+\tensor{\tau}{^{\alpha\gamma\lambda}}\tensor{\tau}{_{\alpha\gamma\lambda}}\right).\label{eq:tau tilde def}
\end{align}
Second, torsion does not propagate since it is nonzero only inside matter, where it mediates a spin-spin contact interaction. Nevertheless, the spin of matter can still influence the spacetime geometry  in vacuum, albeit only indirectly and very weakly, as can be inferred from the structure of Eq. \eqref{eq:field eqs for metric G}, where   the quadratic-in-spin corrections are suppressed by a factor  $\chi$ relative to the other contributions.

\subsection{The linearized regime}\label{Sec:linearized-EC-theory}

The linearized approximation of EC theory applies to the regime where both the metric and torsion  can be
regarded as weak   (see Ref. \cite{arkuszewski1974linearized} and  Sec. IIIC in Ref. \cite{Battista:2021rlh}).   In such a scenario,  these fields can be simultaneously expanded  around the (torsionless) Minkowski background, and  quantities of second order (i.e., involving quadratic contributions in the perturbations or  their derivatives) can be consistently neglected.  

Similarly to GR (see e.g. Refs. \cite{MTW1973,Carroll:2004st,Maggiore:2007ulw}), we introduce a \textit{global nearly Lorentzian coordinate system}  and expand the metric as
\begin{equation}\label{eq:metric perturbation definition}
g_{\mu\nu}=\eta_{\mu\nu}+h_{\mu\nu},
\end{equation}
where  $|h_{\mu\nu}| \ll 1$ and $\eta_{\mu\nu}=\mathrm{diag}(-1,1,1,1)$ is the Minkowski  metric, which we use to raise and lower tensor indices. It is worth noticing that  standard Cartesian coordinates $x^\mu=(ct,x,y,z)$ on the Minkowski background  fall within  the class of coordinates underlying Eq. \eqref{eq:metric perturbation definition},  unlike, for example,  spherical ones.  To fix the ideas, for our calculations in linearized EC gravity we will use coordinates $x^\mu=(ct,x,y,z)$, unless otherwise stated.

As a consequence of  Eq. \eqref{eq:metric perturbation definition},  the inverse metric reads as  $g^{\mu\nu}=\eta^{\mu\nu}-h^{\mu\nu}+\mathrm{O}_g(2)$, and one  thus finds
\begin{align}
\tensor{\hat{\Gamma}}{^\lambda_{\mu\nu}}=&\frac{1}{2}\eta^{\lambda\alpha}\left( \partial_{\mu}h_{\nu\alpha} + \partial_{\nu}h_{\alpha \mu} - \partial_{\alpha}h_{\mu\nu}\right)+\mathrm{O}_g(2),
\end{align}
which, jointly  with condition $\tensor{K}{_{\mu\nu}^\lambda}=\mathrm{O}_g(1)$, leads to 
\begin{align}
 \nabla_{\mu}=&\partial_\mu+\mathrm{O}_g(1).   \label{linearized-nalbla}
\end{align}
Here, we have introduced the notation $\mathrm{O}_{g}(n)$ to indicate terms of order $n$ in the \qm{gravitational} perturbations (e.g., $h^{\rho \sigma}h_{\mu\nu}$,  $h^{\rho \sigma} S_{\mu\nu\lambda}$, $\partial_\alpha h^{\rho \sigma}  S_{\mu\nu \lambda}$ represent  $\mathrm{O}_g(2)$ contributions). 

Owing to above formulas and the relation ${\tilde{\tau}}^{\mu\nu}=\mathrm{O}_g(2)$ (see Eq. \eqref{eq:tau tilde def}),  the EC field equations   \eqref{eq:field eqs for metric G} take the following form in the weak-field limit:
\begin{align}\label{eq:linearized EC equations}
-&\Box\tensor{\bar{h}}{_{\mu\nu}}+\partial_\sigma\left(\partial_\nu\tensor{\bar{h}}{^\sigma_\mu}+\partial_\mu\tensor{\bar{h}}{^\sigma_\nu}-\eta_{\mu\nu}\partial_\lambda\tensor{\bar{h}}{^{\sigma\lambda}}\right)+ \mathrm{O}_g(2)
\nn \\
&=2\chi T_{\mu\nu},
\end{align}
where  $\Box:=\eta^{\mu\nu}\partial_{\mu}\partial_{\nu}$  is  the flat-space d'Alembert operator, 
\begin{equation}
\bar{h}_{\mu\nu}:= h_{\mu\nu}-\frac{1}{2}\eta_{\mu\nu} \tensor{h}{^\alpha_\alpha} \ , \label{eq: def trace-reversed perturbation}
\end{equation}
denotes the trace-reversed perturbation, and 
\begin{equation}
    T_{\mu\nu}=\mathbb{T}_{\mu\nu}+\partial_\lambda\left(-\tensor{\tau}{_{\mu\nu}^\lambda}+\tensor{\tau}{_\nu^\lambda_\mu}-\tensor{\tau}{^\lambda_{\mu\nu}}\right)+\mathrm{O}_g(2),
\end{equation}
which readily stems from Eq. \eqref{bold-T-tensor} and corresponds to the linearized EC version of the Belinfante-Rosenfeld procedure.

It is thus clear that by adopting  the  \textit{Lorentz gauge}  
\begin{equation}\label{eq:Lorentz gauge}
\partial^\nu\tensor{\bar{h}}{_{\mu\nu}}=0 \ , 
\end{equation}
Eq. \eqref{eq:linearized EC equations} boils down, after discarding $\mathrm{O}_g(2)$ corrections,  to the flat-space wave equation 
\begin{equation}\label{eq:linearized EC equations in Lorentz gauge}
    \Box\tensor{\bar{h}}{_{\mu\nu}}=-2\chi T_{\mu\nu} \ .
\end{equation}
The corresponding formal  solution, appropriate for radiative problems, can be  expressed via the retarded Green function as \cite{Maggiore:2007ulw,MTW1973}
\begin{equation}\label{eq:general solution D'alambert equation}
\tensor{\bar{h}}{_{\mu\nu}}(t,\Vec{x})=\frac{4G}{c^4}\int  \frac{\dd^3 x^\prime}{\abs{\Vec{x}-\Vec{x}^\prime}}T_{\mu\nu}\left(t-\frac{\abs{\Vec{x}-\Vec{x}^\prime}}{c},\Vec{x}^\prime\right),
\end{equation}
the combination $t- c^{-1}\abs{\Vec{x}-\Vec{x}^\prime}$ representing the so-called retarded time. 

A final crucial remark is in order. As discussed above, within  the linearized EC approximation    both  the metric $h_{\mu \nu}$ and the torsion $\tensor{S}{_{\mu\nu}^{\lambda}}$ are  \emph{formally} of the same perturbative order $\mathrm{O}_{g}(1)$, which justifies  treating them as comparable for bookkeeping purposes. Despite that,   phenomenologically the situation is different. While $h_{\mu \nu}$ can encompass both classical and quantum contributions, the latter understood   as  involving a factor $\hbar$, $\tensor{S}{_{\mu\nu}^{\lambda}}$ is a \qm{fully quantum} object, being related via Eq. \eqref{eq:field eqs spin R} to the canonical spin tensor.

\section{The exterior geometry  of a slowly rotating  Weyssenhoff-fluid source}\label{sec:Weyssenhoff_fluid_source_and_solution}

The key feature of the \textit{Weyssenhoff fluid}  is that it incorporates the effects of the intrinsic angular momentum of its constituents into the fluid description. As pointed out before, owing to its potentially significant implications in both high-density cosmological and astrophysical contexts, this model is extensively studied  within EC gravity (see e.g. Refs. \cite{Obukhov:1987yu,Boehmer:2006gd,Bohmer2017}). 

In this section, we introduce a Weyssenhoff fluid that  undergoes a macroscopic rotation and thus  carries a classical angular momentum in addition to the microscopic quantum spin. To the best of our knowledge, this combined configuration is derived  here  for the first time  in the literature. 

Specifically, we consider a  \textit{slowly rotating, stationary, axisymmetric, and isolated}  Weyssenhoff source in the weak-field regime, where the linearized EC theory can be reliably invoked.
After outlining the key aspects of the Weyssenhoff paradigm in Sec. \ref{sec:intro-Weyssenhoff}, 
we determine the corresponding  exterior geometry via a two-step program.  First, we work out the explicit form of the metric energy-momentum tensor $T_{\mu \nu}$ in Sec. \ref{sec:rotating-Weyssenhoff}. Second, we  formally solve the ensuing linearized EC equations by means of a multipole expansion  in Sec. \ref{sec:the solution}. We conclude our analysis by addressing the matching conditions  across the fluid  surface in Sec. \ref{sec:junction conditions}.

As we shall see, our model  generalizes the setup of Ref.   \cite{arkuszewski1974linearized}, which deals with the gravitational field generated by a static Weyssenhoff-fluid setup.

\subsection{A brief overview of the Weyssenhoff fluid}\label{sec:intro-Weyssenhoff}

The fluid description of a continuous medium relies on a coarse-grained approximation of a large collection of particles, where molecular fluctuations are smoothed over and each fluid element represents a local average of the matter it contains \cite{Rezzolla,Poisson_Will_2014}.    

The  Weyssenhoff fluid  furnishes a hydrodynamical model of a quantum system. Within this framework,  each fluid element is characterized by  an intrinsic spin density encoded by the tensor 
\begin{equation}\label{eq:def spin tensor smunu from tau}
\tensor{s}{_{\mu\nu}}=-\frac{1}{c^2}\tensor{\tau}{_{\mu\nu}^\lambda}u_\lambda=\tensor{s}{_{[\mu\nu]}}.
\end{equation}
Here,  the  four-velocity $u^\lambda = \dd x^\lambda / \dd \tau$, with $\tau$  the proper time measured by a comoving observer,  obeys  the  normalization relation $u^\mu u_\mu=-c^2$ and can  be arranged as
\begin{equation}\label{eq:four-velocity 2}
u^\mu=\frac{u^0}{c}(c,v^i),
\end{equation}
$v^i= c u^i/u^0 = \dd x^i/\dd t $ being  the coordinate-time three-velocity and, by a slight abuse of terminology,  $u^0/c$ the Lorentz factor \footnote{Owing to the normalization condition $u^\mu u_\mu=-c^2$, the four-velocity  can be arranged in general  as in Eq. \eqref{eq:four-velocity 2}, with $u^0/c=\left(-g_{00}-2g_{0i}v^i/c-g_{ij}v^iv^j/c^2\right)^{-1/2}$. This shows that the normalization factor $u^0/c$ depends not only on $v^i$, but also on the metric components, and is therefore  coordinate dependent.  More generally, the Lorentz factor of the fluid relative to an observer with four-velocity $\mathscr{U}^\mu$ is the scalar $\gamma_{(\mathscr{U})}=-u_\mu \mathscr{U}^\mu/c^2$ (see e.g. Eq. (18) in Ref. \cite{DeFalco2018}). In the nearly Lorentzian coordinates adopted below, the quantity $u^0/c$  reduces, at zeroth order in the gravitational perturbations, to the usual special-relativistic Lorentz factor  (cf. Eq. \eqref{eq:Lorentz-factor-rot}).}.

To correctly account for the three physical degrees of freedom associated with the spin, and to ensure the integrability of the fluid equations of motion, one typically imposes    three  spin supplementary constraints on $s_{\mu \nu}$  \cite{Battista:2021rlh,Costa:2014nta,Battista:2022hmv}. Hereafter, we will adopt the \textit{Frenkel condition}\footnote{Although there exist situations in which the Frenkel condition does not hold (see e.g. Refs.  \cite{Becattini:2007nd,Becattini:2009wh}), it can be safely assumed for  polarized sources, which is  the regime considered in this paper.} \cite{Obukhov:1987yu}
\begin{equation}\label{eq:Frenkel condition}
\tensor{s}{_{\mu\nu}}u^{\nu}=0,
\end{equation}
which covariantly enforces  that $\tensor{s}{_{\mu\nu}}$ is spacelike  in the fluid rest frame;  more details will be provided in Sec.  \ref{sec:more-Frenkel}. 

The Weyssenhoff fluid is characterized by the canonical stress-energy tensor (see Eq. \eqref{bold-T-tensor})
\begin{equation}\label{eq:energy-momentum weyssenhoff}
\tensor{\mathbb{T}}{_{\mu\nu}}=(\rho+p)\frac{u_{\mu}u_{\nu}}{c^2}+pg_{\mu\nu}-\frac{2}{c^2}a^{\beta}\tensor{s}{_{\beta\mu}}u_{\nu},
\end{equation}
where $\rho$ and $p$ are the  total energy density and pressure, respectively, and  $a^\mu=u^\lambda\nabla_{\lambda}u^\mu$  the  four-acceleration. In addition, the tensors $T_{\mu\nu}$ and $\tensor{\tilde{\tau}}{_{\mu\nu}}$  take the  form (cf. Eqs.  \eqref{eq:def dyn canon energy momentum} and \eqref{eq:tau tilde def}) 
\begin{align}
T_{\mu\nu}=&(\rho+p)\frac{u_{\mu}u_{\nu}}{c^2}+pg_{\mu\nu}\nonumber\\
&+2\left(\frac{u_\gamma u^\lambda}{c^2}-\tensor{\delta}{^\lambda_\gamma}\right){\Hat{\nabla}}_\lambda\left(\tensor{s}{^\gamma_{(\mu}}u_{\nu)}\right)\nonumber\\
&-2\chi\left(s^2u_\mu u_\nu+c^2s_{\mu\gamma}\tensor{s}{_\nu^\gamma}\right),\label{eq:symmetric stress energy tensor weyssenhoff}\\
\tensor{\tilde{\tau}}{_{\mu\nu}}=&2c^2s_{\mu\gamma} \tensor{s}{_\nu^\gamma} +s^2 c^2 \left(\frac{u_\mu u_\nu}{c^2}-\frac{1}{2}g_{\mu\nu}\right),
\end{align}
where  $s^2:=\tensor{s}{^{\alpha\beta}}\tensor{s}{_{\alpha\beta}}$ is the (conserved \cite{Obukhov:1987yu,Battista:2022hmv}) spin density scalar.

Combining the above formulas, one obtains from Eq. \eqref{eq:field eqs for metric G} 
\begin{align}\label{eq:EC equations exact Weysshenoff}
{\hat{G}}_{\mu\nu}=&\chi\left[(\rho_{\mathrm{eff}}+p_{\mathrm{eff}})\frac{u_{\mu}u_{\nu}}{c^2}+p_{\mathrm{eff}}g_{\mu\nu}\right.\nonumber \\
&\left.+2\left(\frac{u_\gamma u^\lambda}{c^2}-\tensor{\delta}{^\lambda_\gamma}\right){\Hat{\nabla}}_\lambda\left(\tensor{s}{^\gamma_{(\mu}}u_{\nu)}\right)\right],
\end{align}
with
\begin{subequations}
\begin{align}
\rho_{\mathrm{eff}}&:=\rho-\rho_{\mathrm{sc}},\quad\quad p_{\mathrm{eff}}:=p-p_{\mathrm{sc}},\label{eq:Weysshenoff effective density and pressure exact}
\end{align}
\text{and}
\begin{align}
\rho_{\mathrm{sc}}&\equiv p_{\mathrm{sc}}=\frac{1}{2}\chi c^2s^2.\label{spin-contact correction}
\end{align}
\end{subequations}
EC field equations are thus equivalent to Einstein equations for a perfect fluid with the usual density and pressure replaced by the effective quantities \eqref{eq:Weysshenoff effective density and pressure exact}, along with the gradient term  ${\Hat{\nabla}}_\lambda\left(\tensor{s}{^\gamma_{(\mu}}u_{\nu)}\right)$, which vanishes upon averaging  if the spins of the individual fluid elements are randomly oriented. The contribution introduced by $\rho_{\mathrm{sc}}$ and $p_{\mathrm{sc}}$ amounts to a repulsive force that can have a crucial role in the avoidance of gravitational singularities in collapsing-star models and in the early stages of the universe evolution  \cite{Trautman1973,Poplawski2009,Poplawski2010,Wu2024}. These quadratic-in-spin corrections, which are consistent with the phenomenology  expected from  the Pauli exclusion principle, stem from the peculiar  spin contact interactions of EC gravity.

\subsubsection{More on the Frenkel condition}\label{sec:more-Frenkel}

Let us now   give further details on the physical implications of the Frenkel condition. 

Similarly to the Faraday tensor \cite{MTW1973,Jackson:1998nia},   $s_{\mu\nu}$    admits the \textit{covariant} decomposition
\begin{align}
s_{\mu\nu}=\frac{1}{c}\left(u_{\mu} E_{\nu} - u_{\nu} E_{\mu} \right) + B_{\mu\nu}, \label{eq:spin density in terms of E and B}
\end{align}
where the \qm{electric} four-vector $E_\mu$ and the \qm{magnetic} tensor $B_{\mu\nu}=B_{[\mu \nu]}$ are defined through
\begin{align}
& E_{\nu}=-\frac{1}{c}u^{\mu}s_{\mu\nu},\label{eq:def campo elettrico spin}\\
& B_{\mu\nu}u^{\nu}=0\label{eq:ortogonalita SEM u}.
\end{align}

Since both $E_{\nu}$ and $B_{\mu\nu}$ contain three independent components, a simple  counting argument shows that one could enforce the correct number of spin degrees of freedom by imposing either $E_{\nu}=0$ or $B_{\mu \nu}=0$. The physical reason why the Frenkel condition  \eqref{eq:Frenkel condition} selects the former rather than the latter is that retaining a nonvanishing $B_{\mu \nu}$ permits us to properly define  the Pauli-Lubanski four-vector in a covariant manner
\cite{MTW1973,Speranza:2020ilk} 
\begin{equation}\label{eq:def Pauli-Lubanski}
\ell_\alpha=-\frac{1}{2}\epsilon_{\alpha\beta\mu\nu}\frac{u^\beta}{c} s^{\mu\nu}=-\frac{1}{2}\epsilon_{\alpha\beta\mu\nu}\frac{u^\beta}{c} B^{\mu\nu},
\end{equation}
where $\epsilon_{\alpha\beta\mu\nu}$ is the four-dimensional totally antisymmetric Levi-Civita symbol. Thus, $\ell_\alpha $ has the essential features to embody a truly covariant representation of the spin density vector: (i) it  carries three independent degrees of freedom, as  $\ell_\alpha u^\alpha=0$ by construction; (ii) it transforms as an axial vector. 

This identification can be  clarified as follows. If we introduce the  three-vectors \cite{Boehmer:2006gd,Battista:2021rlh,Battista:2022hmv} 
\begin{align}
\vec{s}=&(s^1,s^2,s^3)\equiv(s^{23},s^{31},s^{12}),\label{eq:magnetic components of spin}   \\
\vec{q}=&(q^1,q^2,q^3)\equiv(s^{10},s^{20},s^{30}), \label{eq:electric components of spin}
\end{align}
and exploit Eq. \eqref{eq:four-velocity 2}, the Frenkel condition \eqref{eq:Frenkel condition}  can be recast as
\begin{subequations}
\label{Frenkel-explicit-comp}
    \begin{align}
    s_{i0}&=-s_{ij}\frac{v^j}{c},\label{eq:Frenkel i0}
    \\
    s_{0i}v^i&=0.\label{eq:Frenkel i0-2}
    \end{align}
\end{subequations}
Adopting nearly Lorentzian coordinates, we can expand  identity \eqref{eq:Frenkel i0} as
\begin{equation}\label{q-equals-v-times-s}
    \vec{q}=\frac{1}{c}\vec{v}\times\vec{s}+\mathrm{O}_g(2),
\end{equation}
where we have  used  the  formulas
\begin{align}
s^i&=\frac{1}{2}\epsilon^{ilm}s_{lm}+\mathrm{O}_g(2),
\nonumber \\
s_{ij}&=\epsilon_{lij}s^{l}+\mathrm{O}_g(2).\label{eq:linearized spin vector from spin tensor}
\end{align}

Relation \eqref{q-equals-v-times-s} closely mirrors the Lorentz-transformation law connecting the electric and magnetic fields \cite{Resnick-book1998}, which naturally suggests interpreting  $\vec{q}$ and $\vec{s}$ as the electric and magnetic components of $s_{\mu \nu}$. This analogy becomes clearer in the (hypothetical)  global rest frame of the fluid, where we use the  symbol \qm{$\doteq$} to denote the corresponding identities. Here,  $E_\mu$, $B_{\mu\nu}$, and $\ell_{\mu}$ are purely spatial, $u^\mu\doteq (c,0,0,0)$, and Eqs. \eqref{eq:magnetic components of spin} and \eqref{eq:electric components of spin} yield 
\begin{align}
\vec{s} \doteq &(B^{23},B^{31},B^{12})\doteq \vec{\ell}  +\mathrm{O}_g(2),\label{eq:magnetic components of spin-2}\\
\vec{q} \doteq &(E^1,E^2,E^3). \label{eq:electric components of spin-2}
\end{align}
These  formulas show that  $\vec{s}$ is identified with the spatial components of the Pauli-Lubanski vector, while the electric piece $\vec{q}\doteq 0$ owing to Eq. \eqref{eq:Frenkel condition}.  However,  the global frame underlying the above calculations does  not exist in general, and the Frenkel condition guarantees that $\vec{q}$  vanishes only \textit{pointwise} in  the instantaneous rest frame of each  fluid element, but  not in an arbitrary coordinate system. This observation is  crucial, as we will see in Sec. \ref{sec:rotating-Weyssenhoff} that  the time-time component of the stress-energy tensor of a rotating Weyssenhoff fluid exhibits a peculiar (spin-orbit) correction due precisely to $\vec{q}$.

\subsection{The stress-energy tensor }\label{sec:rotating-Weyssenhoff}

The first step in  deriving the gravitational field produced by  the rotating Weyssenhoff fluid is to compute the corresponding metric stress-energy tensor $T_{\mu \nu}$, which enters the right-hand side of the linearized EC field equations \eqref{eq:linearized EC equations in Lorentz gauge}.

It  follows from  Eq. \eqref{eq:symmetric stress energy tensor weyssenhoff} that,  in the weak-field approximation, $T_{\mu \nu}$ is given by
\begin{align}
    T_{\mu\nu}=&(\rho+p)\frac{u_{\mu}u_{\nu}}{c^2}+p\eta_{\mu\nu}\nonumber\\
    &+2\left(\frac{u_\gamma u^\lambda}{c^2}-\tensor{\delta}{^\lambda_\gamma}\right)\partial_\lambda\left(\tensor{s}{^\gamma_{(\mu}}u_{\nu)}\right) +\mathrm{O}_{g}(2) \ . \label{eq:linearized symmetric stress energy tensor weyssenhoff}
\end{align}
At this stage, however, the above expression remains completely general. To specialize it to our configuration,  it must  be adapted to the assumed  properties of the source. We investigate the constraints arising from  stationarity and  axial  symmetry  in Sec. \ref{Sec:symmetries-source}, while the  slow-rotation hypothesis is introduced in Sec. \ref{Sec:Slow-rotation-approximation}, where we also work out the components  of $T_{\mu\nu}$ explicitly. As we will see, among the new velocity-related corrections, a novel  spin-orbit coupling emerges that  modifies the  energy density  of the source. We  inspect this  contribution in detail in Sec. \ref{Sec:estimate-spin-orbit}.

\subsubsection{The hypotheses of  stationarity and axial  symmetry }\label{Sec:symmetries-source}

It is clear from Eq. \eqref{eq:linearized symmetric stress energy tensor weyssenhoff}   that $T_{\mu \nu}$ depends  on four variables:  the energy density, pressure, four-velocity and spin density. Since $\rho$ and $p$ are scalar fields,  stationarity and axial  symmetry simply require  them to  be  time-independent and constant along circles centered on the symmetry axis. By contrast, the situation is slightly more involved for $u^\mu$ and $s_{\mu\nu}$, as axial symmetry imposes  additional restrictions on their components.

Let us first deal with the four-velocity. Differently from  the static configuration  \cite{arkuszewski1974linearized},  the presence of a  macroscopic rotation entails that  no  (asymptotically inertial) nearly Lorentzian frame can be defined where the fluid is globally at rest (see our discussion below Eq. \eqref{eq:linearized spin vector from spin tensor}). Consequently,   $u^\alpha(x)\neq(c,0,0,0)$, except possibly on the symmetry axis. Had a coordinate system  existed where $u^\alpha=(c,0,0,0)$ everywhere,  it  would necessarily have been rotating relative to a nearly Lorentzian frame and thus  would not itself  qualify as nearly Lorentzian.

Given the above premises, we can obtain the explicit form of  the four-velocity in a straightforward way if we first adopt a set of spherical coordinates  $\tilde{x}^\mu=(ct,r,\theta, \varphi)$ (the specific  definition of $r$ is not essential in the linearized regime).  Supposing hereafter that the symmetry axis coincides with the $z$-direction, a  purely rotational (i.e., circular)  stationary, axisymmetric source admits the general formula   (cf. Eq. (11.145) in Ref. \cite{Rezzolla}): 
\begin{align}\label{four-velocity-spherical-aux}
\tilde{u}^{\mu} \left(\tilde{x}\right)=&\frac{\dd \tilde{x}^\mu}{\dd \tau}=(\tilde{u}^0,\tilde{u}^r,\tilde{u}^\theta,\tilde{u}^\varphi)
\nonumber \\
=&\frac{\tilde{u}^0\left(r,\theta\right)}{c}\bigl(c, 0, 0, \omega(r,\theta)\bigr),
\end{align}
where the time-independent function $\omega(r,\theta)$      physically represents the (counter-clockwise) rotation rate of the fluid at the spacetime points specified by  $r$ and $\theta$, as measured by an observer at infinity (cf. Eq. (11.146) in Ref. \cite{Rezzolla}):
\begin{equation}
    \dv{\varphi}{t}=\dv{\varphi}{\tau}\left(\dv{t}{\tau}\right)^{-1}=c\frac{\tilde{u}^\varphi}{\tilde{u}^0}=\omega(r,\theta).
\end{equation}
The  exact profile of $\omega$ depends both on the properties of the fluid (e.g.,  whether it undergoes rigid or differential rotation) as well as  on the precise representation of the variables $r$ and $\theta$.

Transforming expression \eqref{four-velocity-spherical-aux} from spherical coordinates  $\tilde{x}^\mu$ back  to the nearly Lorentzian Cartesian system $x^\mu=(ct,x,y,z)$, and using Eq. \eqref{eq:four-velocity 2}, the four-velocity takes the form 
\begin{equation}\label{eq:four-velocity 3}
 u^{\mu}(x)=(u^0,u^x,u^y,u^z)=\frac{u^0}{c}\bigl(c, -\omega(r,\theta) y,  \omega(r,\theta) x,0\bigr). 
\end{equation}
Therefore, the coordinate three-velocity reads as 
\begin{equation}
    \vec{v}=\vec{\omega}\times\vec{x}=  (-\omega y,\omega x, 0)=(v^x,v^y,v^z) \ ,\label{eq:v-omega-times-x}
\end{equation}
and the Lorentz factor becomes
\begin{align}
\frac{u^0}{c}=&\left(1-\frac{\abs{\vec{v}}^2}{c^2}\right)^{-1/2}+\mathrm{O}_g(1)
\nonumber \\
=&\left(1-\frac{\omega^2\varrho^2}{c^2}\right)^{-1/2}+\mathrm{O}_{g}(1),
\label{eq:Lorentz-factor-rot}
\end{align}
with $\vec{\omega}\left(r,\theta\right)=\omega\left(r,\theta\right)\hat{z}$ and $\varrho:=\sqrt{x^2+y^2}$.

We now examine the spin density. Consistently with the axial symmetry, we  assume that the Weyssenhoff fluid has spins everywhere aligned with the $z$-axis, i.e., $s^x=s^y=0$ (cf. Eq. \eqref{eq:magnetic components of spin}). In this setup,  it is natural to impose the Frenkel condition \eqref{eq:Frenkel condition}, which, in view of Eqs. \eqref{eq:linearized spin vector from spin tensor} and  \eqref{eq:four-velocity 3}, yields 
\begin{subequations}
\label{relations-Frenkel}
\begin{align}
s_{x0}+s^zx\frac{\omega}{c}+\mathrm{O}_{g}(2)&=0,\label{Frenkel-1}\\
s_{y0}+s^zy\frac{\omega}{c}+\mathrm{O}_{g}(2)&=0,\label{Frenkel-2}\\
\omega (s_{y0} x-s_{x0}y)&=0,\label{Frenkel-3} \\
s_{z0} u^0 +\mathrm{O}_{g}(2)&=0. \label{Frenkel-4}
\end{align}
\end{subequations}
Discarding second-order  gravitational corrections, one   immediately sees from Eq. \eqref{Frenkel-4} that $s_{z0}=0$. Hence, the only nonvanishing entries of the spin density tensor are  $ s_{x0}$,  $s_{y0}$, and $s_{xy}\equiv s^z$, the latter being the sole independent component. This result can be easily verified. In fact,  Eq. \eqref{Frenkel-3} imposes no additional constraint on $s_{\mu \nu}$, as it is  automatically satisfied once Eqs. \eqref{Frenkel-1} and \eqref{Frenkel-2} hold. Once $s^z$ and $\omega$ are specified, Eqs. \eqref{Frenkel-1} and \eqref{Frenkel-2} fully determine $s_{x0}$ and $s_{y0}$, which continue to vary  through  their dependence  on the coordinates $x$ and $y$ even in the particular scenario where $s^z$ and/or $\omega$ are constant. Furthermore, when $\omega\neq0$, Eq. \eqref{Frenkel-3} gives rise  to the  geometric relation
\begin{equation}
    \frac{s_{y0}}{s_{x0}}=\tan\varphi,\quad {\rm for}\,\,\,x\neq0,
\end{equation}
which is the expected consequence of the axial symmetry of the configuration. 

A further crucial outcome of Eq. \eqref{relations-Frenkel} is that  the electric part $\Vec{q}=\left(s^{x0}, s^{y0}, s^{z0}\right)$ of the spin (see Eq. \eqref{eq:electric components of spin}) is in general nonvanishing whenever  $\omega\neq 0$, i.e., for a rotating fluid. As pointed out  before,  this  is perfectly compatible  with the Frenkel condition.

\subsubsection{The slow-rotation approximation and structure of $T_{\mu \nu}$}\label{Sec:Slow-rotation-approximation}

Having seen in the previous section how the source variables entering $T_{\mu \nu}$ are affected by the requirements of stationarity and axial symmetry,  we can now introduce the additional slow-rotation hypothesis. 

Accordingly, we assume that the \qm{velocity-related} parameters $\vec{\omega}$ and $ \vec{v} $ be  $\mathrm{O}_{\textit{v}}(1)$ quantities,  with $\mathrm{O}_{v}(n) \equiv \mathrm{O}\left(\vert \vec{v} \vert^n\right) $, while, like before, $h_{\mu\nu}$, $s_{\mu\nu}$, $\rho$, and $p$ are regarded as  $\mathrm{O}_{\textit{g}}(1)$ contributions.  Since  the linearized EC theory (see Sec. \ref{Sec:linearized-EC-theory})  does not address  the full regime of strongly self-gravitating sources,  the   weak-field and low-velocity expansions can be treated as  independent. Therefore, in this setup we  retain corrections of order $\mathrm{O}_g(1)$, $\mathrm{O}_{v}(1) $, as well as their products $\mathrm{O}_g(1) \mathrm{O}_v(1)$,  while  safely discarding $\mathrm{O}_g(2)$, $\mathrm{O}_{v}(2)$, and higher. This scheme is consistent with the structure of $s_{\mu\nu}$, for which  we initially take  $s_{\mu \nu}=\mathrm{O}_{g}(1)$, and  the Frenkel condition then refines this to  $s_{0i}=\mathrm{O}_{g}(1) \mathrm{O}_{v}(1)$ and  $s_{ij}=\mathrm{O}_{g}(1) $ (see Eqs. \eqref{Frenkel-explicit-comp} and \eqref{relations-Frenkel}).

Including mixed factors  $ \mathrm{O}_{g}(1) \mathrm{O}_{v}(1) $ is  standard in linearized GR and is precisely what allows one to recover  the exterior field of a weakly gravitating,   rotating system (see Sec. 19.1 in Ref. \cite{MTW1973}). Our approach extends this procedure, as  keeping factors like  $v^j s_{ij} \sim \mathrm{O}_{v}(1)  \mathrm{O}_{g}(1) $  amounts to retaining $ \mathrm{O}(v)\mathrm{O}(\hbar) $ terms that capture spin-rotation couplings (recall that $s_{\mu \nu}$ is  intrinsically $\hbar$-dependent).  This treatment aligns with the spirit of  the static-source limit analyzed in Ref.  \cite{arkuszewski1974linearized}.

Under the hypothesis of slow rotation, we find that the linearized stress-energy tensor \eqref{eq:linearized symmetric stress energy tensor weyssenhoff} is (for simplicity, the  remainders  are omitted hereafter)\footnote{Notice that Eq. \eqref{eq:stress energy tensor cartesian} is written in a form that is explicitly invariant under spatial rotations.} 
\begin{subequations}\label{eq:stress energy tensor cartesian}
    \begin{align}
    T_{00}=&\rho+2c\partial_i\tensor{s}{_{i0}},\label{eq:stress energy tensor cartesian spin e angular 00}\\
    T_{0j}=&-(\rho+p)\frac{v^j}{c}+c\partial_is_{ij},\label{eq:stress energy tensor cartesian spin e angular 0j}\\
    T_{ij}=&p\delta_{ij}-\partial_l\left(\tensor{s}{_{li}}v^j\right)-\partial_l\left(\tensor{s}{_{lj}}v^i\right),\label{eq:stress energy tensor cartesian spin e angular ij}
    \end{align}
\end{subequations}
which, in terms of the  electric and magnetic spin three-vectors $\vec{q}$ and $\vec{s}$, becomes 
\begin{subequations}\label{eq:T from q and s}
    \begin{align}
    T_{00}=&\rho-2c\vec{\nabla}\cdot\vec{q},\label{eq:T00 in terms of q and s}\\
    T_{0j}=&-(\rho+p)\frac{v^j}{c}-c\left(\vec{\nabla}\times\vec{s}\right)^j,\label{eq:T0j in terms of q and s}\\
T_{ij}=&p\delta_{ij}+2\left(\vec{v}\otimes\vec{\nabla}\times\vec{s}\right)^{(ij)}-2\left(\vec{s}\times\vec{\nabla}\right){}^{(i}v^{j)}. \label{eq:Tij in terms of q and s}
\end{align}
\end{subequations}
Here,  the expression for $T_{0j}$  takes into account that $s_{i0}=\mathrm{O}_{g}(1)\mathrm{O}_{v}(1)$ and  \qm{$\otimes$} is the ordinary tensor product. 

Owing to Eqs.    \eqref{eq:magnetic components of spin} and \eqref{eq:v-omega-times-x},   in the Cartesian coordinates $x^{\mu}=(ct,x,y,z)$, $T_{\mu \nu}$  explicitly reads  
\begin{subequations}\label{eq:stress-energy-explicit}
\begin{align}
    T_{00}=&\rho-2\left[2s^z\omega+x\partial_x(s^z\omega)+y\partial_y(s^z\omega)\right],\label{T-00-explicit-1}\\
    T_{0x}=&(\rho+p)\frac{\omega y}{c}
    -c\partial_ys^{z},\\
    T_{0y}=&-(\rho+p)\frac{\omega x}{c}
    +c\partial_xs^{z},\\
    T_{xx}=&p-2\partial_y(ys^z\omega),\\
    T_{xy}=&y\partial_x(s^z\omega)+x\partial_y(s^z\omega),\\
    T_{yy}=&p-2\partial_x(xs^z\omega),\\
    T_{zz}=&p.
\end{align}
\end{subequations}
As expected,  the above formulas boil down to those pertaining to   nonrotating sources \cite{arkuszewski1974linearized}. In contrast, when  $\omega \neq 0$, $T_{\mu \nu}$ acquires  additional velocity-dependent corrections. Specifically,  the diagonal elements $T_{xx}$ and $T_{yy}$ modify the effective pressure in an anisotropic way,  the mixed   $T_{0j}$ encode, as we will see,   the macroscopic angular momentum of the source (cf. Eqs. \eqref{eq:final metric 0i} and \eqref{eq:total macro spin of the source} below), and the off-diagonal term $T_{xy}$  describes the $xy$-shear stress. 

A separate discussion concerns $ T_{00}$. It  can be naturally  recast as 
\begin{align}
 T_{00} =     \rho -2 \rho_{\text{so}}, 
 \label{T-00-expr-so-density}
\end{align}
where we have introduced 
\begin{align}
\rho_{\text{so}}:=&c\vec{\nabla}\cdot\vec{q}=-c\partial_i s_{i0}=\partial_i(s_{ij}v^j)
\nonumber \\
=&2s^z\omega+x\partial_x(s^z\omega)+y\partial_y(s^z\omega) \ ,
\label{eq:effective spin mass density}
\end{align}
the last equality on the first line stemming from Eq. \eqref{eq:Frenkel i0}.  This new  contribution is nonvanishing only when both  the \textit{macroscopic} angular velocity  $\vec{\omega}$  and the \textit{microscopic} quantum spin $\vec{s}$ are nonzero, and thus  behaves as an effective \qm{spin-orbit energy density}. Being  proportional to  $\vec{\nabla}\cdot\vec{q}$, it is reminiscent of the charge density in  electromagnetism (which  involves the divergence of the electric field), and, likewise, can attain either positive or negative values.  Assuming as a first approximation that  $s^z\omega=\mathrm{constant}$ (as for a uniformly polarized rigid body), the  sign of $\rho_{\text{so}}$  is  determined solely by the relative orientation of $\vec{\omega}$ and $\vec{s}$, as it is positive (resp. negative) when they are aligned (resp. antialigned). This observation  points to a clear spin-orbit interpretation: the   configuration having aligned $\vec{\omega}$ and $\vec{s}$ is  energetically favored, since  it reduces the effective energy density   $\rho -2 \rho_{\text{so}}$ occurring in Eq. \eqref{T-00-expr-so-density}. 

Interestingly, the dependence of $\rho_{\text{so}}$ on the electric vector $\vec{q}$ allows another parallel with Maxwell theory. As  explained before, $\vec{q} \neq 0$ in the nearly Lorentzian frame where the source rotates with  angular velocity $\omega$, while the Frenkel condition  enforces $\vec{q}=0$ locally in each fluid element rest frame. This behavior resembles that of  the electric field generated by a rotating magnet \cite{Backus1956}: the field is absent  in the magnet rest frame but  appears in a rotating one. This frame effect arises from the  intertwining of the electric and magnetic fields under Lorentz transformations  \cite{Resnick-book1998}, and further supports  the  interpretation of Eq. \eqref{q-equals-v-times-s} as providing an electromagnetic analogue for $\vec{q}$ and $\vec{s}$.

\subsubsection{Estimate of the spin-orbit energy density $\rho_{\text{so}}$}\label{Sec:estimate-spin-orbit}

Given its central role in our analysis,  in this  section  we take a closer look at the spin-orbit energy density $\rho_{\text{so}}$. 

Like before, we assume $s^z\omega=\mathrm{constant}$ for simplicity. In this setting,  $\rho_{\text{so}}=2 s^z \omega$ and  $T_{\mu \nu}$ becomes diagonal (cf. Eqs. \eqref{eq:effective spin mass density} and \eqref{eq:stress-energy-explicit})
\begin{subequations}
\begin{align}
   T_{00}=& \rho-4s^z\omega \ ,\\
    T_{xx}=& T_{yy}= p-2s^z \omega \ ,\\
    T_{zz}=&p \ , 
\end{align}
\end{subequations}
showing that the pressure orthogonal to the rotation axis is  reduced by half the factor that affects the energy density.

Assuming a  polarized source consisting of spin-$\tfrac{1}{2}$ particles (e.g. neutrons or electrons), the spin density  can be  modeled by  $s^z=n \hbar/2$, while the rest energy density is  $\rho=mc^2 n$, where $n$ is the particle number density and $m$  the mass of a single constituent \cite{arkuszewski1974linearized,Hehl:1976kj}. Combining  these relations,  yields 
\begin{align}
    \frac{\rho_{\text{so}}}{\rho}\sim\frac{\lambdabar_C}{c}\omega,\label{ratio so}
\end{align}
or, equivalently,
\begin{align}
    \frac{\rho_{\text{so}}}{\rho}\sim\varepsilon \frac{\lambdabar_C}{\mathcal{R}}, 
    \label{ratio so 2}
\end{align}
where $\lambdabar_C=\hbar/\left(m c\right)$ is the reduced Compton wavelength,  $\mathcal{R}$  the characteristic radius of the source, and $\varepsilon:=\omega \mathcal{R}/c$  measures how close the equatorial (tangential)  speed is  to the relativistic limit.

Remarkably, the ratio $\rho_{\text{so}}/\rho$ is independent of  $n$ and hence  the spin-orbit correction can, in principle, be  relevant  even   for gravitational systems of comparatively low density. As expected, this contribution   becomes more important as  $\omega$ increases, and is especially significant for small, rapidly rotating objects. However, $\rho_{\text{so}}/\rho$  cannot be arbitrarily large, for at least two reasons: (i)  causality demands $\varepsilon<1$; (ii) for a noncollapsed source,  $\mathcal{R}$ must exceed some critical value (e.g., in GR a static, spherically symmetric object must have $\mathcal{R}$  larger than the Schwarzschild radius). These physical constraints limiting  $\rho_{\text{so}}/\rho$ actually favor the self-consistency of our approach, since they help the system  remain  within the  slow-rotation regime discussed before.

It is appropriate to make a comparison with the characteristic spin-contact density $\rho_{\mathrm{sc}}$ (see Eq.  \eqref{spin-contact correction}), which we have neglected in the linearized approach owing to its dependence on the spin density scalar  $s^2$.  In this case,  one finds
\begin{align}
\frac{\rho_{\text{sc}}}{\rho}\sim \pi \ell_P^2 \lambdabar_C n\,,\label{ratio sc}
\end{align}
with $\ell_P=\sqrt{\hbar G/c^{3}}$  the Planck length. Differently from Eq. \eqref{ratio so},  the spin-contact term scales linearly with $n$, meaning that it is significant only at extremely  high densities comparable to or  beyond the cutoff   $n_{\rm cut}\sim 1/(\ell_P^2 \lambdabar_C)$ (cf. Eq. (5.4) in Ref. \cite{Hehl:1976kj}). 

The corrections related to $\rho_{\text{so}}$ dominate over the spin-contact ones in  realistic settings, such as astrophysical scenarios comprising neutron stars. To illustrate this, let us consider  the fastest-rotating neutron star  observed to date, identified as  PSR J1748-2446ad  and with  reference values   $\varepsilon \sim 1/4$ and $\mathcal{R} \sim 16\, \text{km}$ \cite{Hessels:2006ze}. Assuming that it is  made up   almost exclusively of neutrons,  for which $\lambdabar_C\sim2.1\times 10^{-16}$ m,  we obtain
\begin{align}
    \frac{\rho_{\text{so}}}{\rho}\sim3\times10^{-21}, \label{ratio-rho-so}
\end{align}
which is roughly twenty orders of magnitude larger  than the spin-contact  ratio \eqref{ratio sc}:
\begin{align}
    \frac{\rho_{\text{sc}}}{\rho}\sim5\times10^{-41},
\end{align}
where we have used the representative  number density for neutron-star matter $n\sim 3\times10^{44}\,{\rm m}^{-3}$ (corresponding to roughly twice the nuclear saturation density \cite{Huth2021,Sedrakian2022}).

Remarkably, the above conclusions remain valid over a broad range of parameters.  For instance,  assuming again an effectively polarized neutron star  and   reducing  $\varepsilon$ by  one order of magnitude, i.e., taking $\varepsilon \sim 10^{-2}$ rather than $\varepsilon \sim 1/4$, while keeping the values of all other parameters the same as before,  still permits recovering the hierarchy   $\rho_{\text{so}}/\rho \gg \rho_{\text{sc}}/\rho$. Indeed,  Eq. \eqref{ratio so 2}  yields in this case $\rho_{\text{so}}/\rho \sim 1.3 \times 10^{-22}$, while   $\rho_{\text{sc}}/\rho $ remains unchanged.

A last remark is in order.   Within the exact treatment of the rotating Weyssenhoff fluid,  we reasonably  expect that $T_{\mu \nu}$ will involve  both $\rho_{\mathrm{so}}$ and  $\rho_{\mathrm{sc}}$. Accordingly, we might define an extended effective energy density  $\rho^\prime_{\mathrm{eff}}=\rho - \rho_{\mathrm{sc}}- 2 \rho_{\mathrm{so}} \equiv \rho_{\mathrm{eff}}-2 \rho_{\mathrm{so}}$ along the lines of Eq. \eqref{eq:Weysshenoff effective density and pressure exact}. This modified  density could have implications in the study of gravitational collapse and singularity avoidance, in analogy with the nonrotating Weyssenhoff scenario, see e.g. Refs. \cite{Poplawski2010,Choudhury2024,Wu2024}.

\subsection{Multipole expansion of the  metric}\label{sec:the solution}

With the explicit expression of the (linearized) stress-energy tensor for a slowly rotating and axisymmetric  Weyssenhoff fluid  at hand (see Eqs. \eqref{eq:stress energy tensor cartesian}-\eqref{eq:stress-energy-explicit}), we can now address the second step of our computational program and  study the resulting exterior geometry. 

As set out earlier, in the Lorentz gauge \eqref{eq:Lorentz gauge}  the  linearized EC equations take the form \eqref{eq:linearized EC equations in Lorentz gauge}. These attain a simplified form in our setup, because stationary implies that the flat-space d'Alembert operator    boils down to the flat-space Laplace operator
\begin{equation}
    \Box\tensor{\bar{h}}{_{\mu\nu}}=\eta^{ij}\partial_{i}\partial_{j}\bar{h}_{\mu\nu}=:\Delta \bar{h}_{\mu\nu},\label{Laplace-flat}
\end{equation}
since $\eta^{00}\partial_{0}\partial_{0}\bar{h}_{\mu\nu}=0$.

Since  the appropriate boundary conditions for an isolated source require that the metric perturbations $\tensor{\bar{h}}{_{\mu\nu}}$ vanish at spatial infinity,  the formal solution  of Eq. \eqref{eq:linearized EC equations in Lorentz gauge} in the limit \eqref{Laplace-flat}   involves the Poisson integral
\begin{equation}\label{eq: h bar in terms of T for a stationary spacetime}
    \tensor{\bar{h}}{_{\mu\nu}}(\Vec{x})=\frac{4G}{c^4}\int \frac{\dd^3 x^\prime}{\abs{\Vec{x}-\Vec{x}^\prime}}T_{\mu\nu}(\Vec{x}^\prime),
\end{equation}
rather than   the retarded Green function   \eqref{eq:general solution D'alambert equation}. 

In line with the fact that  we are concerned  with the geometry \textit{outside} the fluid,  the evaluation of the above integral   can be tackled by means of a multipole expansion, where,  differently from the static case \cite{arkuszewski1974linearized}, we   allow for a nonzero pressure. Following the  approach outlined in  Ref. \cite{MTW1973} (see Eqs. (19.5) and (19.22) therein),  we exploit the gauge freedom to perform a coordinate transformation to a new set of nearly Lorentzian coordinates $x^{\prime \mu}=(ct,\Vec{x}^\prime)$, with $\Vec{x}^\prime = (x',y',z')$,  in which the leading terms of $\bar{h}^{\prime}_{00}$, $\bar{h}^{\prime}_{0i}$, and $\bar{h}^{\prime}_{ij}$ scale as  $\mathrm{O}(1/r^\prime)$, $\mathrm{O}(1/r^{\prime 2})$, and $\mathrm{O}(1/r^{\prime 3})$, respectively (with $r':=\vert \Vec{x}' \vert $). Dropping the primes for simplicity, and using the definition \eqref{eq: def trace-reversed perturbation},   we find that the components of $h_{\mu \nu}$ in the new frame read as,  to the leading nontrivial multipole order, 
\begin{subequations}\label{eq:final metric total}
\begin{align}
    h_{00}=&\frac{2G}{c^2}\frac{M-2M_s}{r} \ ,\label{eq:final metric 00}\\
    h_{0i}=&\frac{2G}{c^3}\frac{\left[\Vec{x}\times(\Vec{J}+2\Vec{S})\right]_i}{r^3} \ ,\label{eq:final metric 0i}\\
    h_{ij}=&\frac{2G}{c^2}\frac{M-2M_s}{r}\delta_{ij} \ ,\label{eq:final metric ij}
\end{align}
\end{subequations}
with  
\begin{subequations}\label{eq:total parameters}
    \begin{align}
    M:=&\frac{1}{c^2}\int\dd^3 x \, \rho(\Vec{x}) \ ,\label{eq:total regular mass of the source}\\
    \Vec{J}:=&\frac{1}{c^2}\int\dd^3x\, (\rho+p)\left(\Vec{x}\times\Vec{v}\right) \ ,\label{eq:total macro spin of the source}\\
     M_s:=&\frac{1}{c^2}\int\dd^3 x \, \rho_{\text{so}}(\Vec{x}) \ ,\label{eq:total spin mass of the source}\\
\Vec{S}:=&\int\dd^3 x \, \Vec{s}(\Vec{x}) \ .\label{eq:total micro spin of the source}
   \end{align}
\end{subequations}

The quantities $M$ and $\Vec{J}$ are  the usual mass and classical angular momentum of the source, respectively  \cite{MTW1973}. The term  $M_s$ represents an
additional effective mass  arising from  the spin-orbit energy density  \eqref{eq:effective spin mass density} and hence linear in both the fluid spin and angular velocity, while $\Vec{S}$ denotes the microscopic  quantum spin of the matter distribution, obtained as the volume integral of the magnetic vector $\Vec{s}$  (see Eq. \eqref{eq:magnetic components of spin}).

The solution \eqref{eq:final metric total} can be identified with a generalized  Lense-Thirring geometry \cite{MTW1973}, where, remarkably, the Newtonian and gravitomagnetic potentials can be obtained through the replacements
\begin{align}
 M \rightarrow M-2M_s, \qquad   \Vec{J} \rightarrow \Vec{J} +2 \vec{S}. \label{replacement-param}
\end{align}
In other words,  the spin-induced modifications affect both sectors of the weak-field geometry. The gravitoelectric correction is encoded in the effective mass $M-2M_s$, whereas the gravitomagnetic field is sourced by the effective angular momentum $\vec J+2\vec S$.  We will exploit such result in Sec. \ref{sec:classical test gyros}. 

Before proceeding further, the following points should be highlighted. First, the quantum spin $\Vec{S}$ enters $h_{0i}$   with a coefficient twice that of the  classical   $\Vec{J}$, and an analogous structure also appears in $h_{00}$ and $h_{ij}$ for the mass terms. Remarkably, this feature  is reminiscent of the normalization factor found in the  analysis of binary dynamics in EC theory, which has been carried out, at the first post-Newtonian level, in Refs. \cite{Battista:2022sci,Battista:2023znv,DeFalco:2023djo}  by applying the point-particle procedure to the Weyssenhoff fluid and within the Frenkel description. Although the two results arise in different settings, they both give precious indications of how the  combined effect of the spin and macroscopic rotation affects (weak) gravitational phenomena within the Weyssenhoff-fluid scenario (and for the spin-current convention adopted in Eq. \eqref{eq: def dyn tau 2}).

Second, it is interesting to note that under the same assumptions as in Sec. \ref{Sec:estimate-spin-orbit}, the neutron star  PSR J1748-2446ad has a ratio  $S/J$  of the same order  as the dimensionless parameter controlling the spin-orbit correction $\rho_{\text{so}}/\rho$ (see Eq. \eqref{ratio-rho-so}):
\begin{align}
    \frac{S}{J}=\frac{5}{4}\frac{\lambdabar}{\varepsilon \mathcal{R}}\sim 4\times 10^{-21}\,. 
\end{align}
This indicates that the spin-orbit contribution to the gravitoelectric sector and the intrinsic-spin correction to the gravitomagnetic field become relevant at comparable scales. 

Last, it is clear that  both $\Vec{J}$ and $M_s$ vanish in the static limit. A naive application of the divergence theorem to Eqs.  \eqref{eq:effective spin mass density} and \eqref{eq:total spin mass of the source} might  suggest that $M_s$ should be zero also in the rotating configuration. However, this need not be the case in general, since $\vec{q}$ can fail to be regular on the boundary of the fluid, as we will discuss in  Sec. \ref{sec:junction conditions}.

\subsection{Junction conditions}\label{sec:junction conditions}

In many gravitational applications,  one must  tackle the problem of matching  two metrics smoothly across a hypersurface $\Sigma$ that divides the spacetime into two regions.  In such situations, the field equations are understood in a distributional sense, and singular terms are either removed or given a proper physical interpretation by imposing so-called \textit{junction conditions}, which configure as constraints on the jumps  of geometric  quantities across $\Sigma$.  In GR, this procedure leads to the well-known Israel formalism and the possible appearance of thin shells \cite{Poisson2009-book}. In EC theory, however,  torsion and spin introduce additional requirements \cite{Arkuszewski:1975fz}, which   do not, in general, guarantee the continuity of the spin density tensor across $\Sigma $. When applied to our model, this implies that the divergence theorem cannot be invoked to  evaluate the spin-orbit mass  $M_s$, which is thus allowed to give a nonzero contribution to  the exterior metric \eqref{eq:final metric total}. 

We  briefly outline the  formalism of junction conditions within EC gravity \cite{Arkuszewski:1975fz} in Sec. \ref{Sec:junction-general}, while  the  application to our setup is discussed in Sec. \ref{Sec:junction-our-setup}.

\subsubsection{The general setup}\label{Sec:junction-general}
%\chi^{ARKUSEWSKI}_{ijk} = -K^{mine}_{kji} (with k,j,i becoming Greek indices in our conventions)

To fix the ideas, let $\Sigma$ be a timelike  hypersurface  described  locally by $y^1=0$ in a suitable coordinate system $y^{\alpha}=(y^0,y^1,y^2,y^3)$ defined on  the Riemann-Cartan spacetime   $\mathcal{M}$.  We say that a function $f:\mathcal{M}\longrightarrow \mathbb{R}$ is of class $C^{-1}_q$ if it is of class $C^q$ in $\mathcal{M}/\Sigma$, and if its derivatives up to (and including) order $q$  admit finite left and right limits on $\Sigma$, uniformly approached from both sides. If, in addition, $f$ is  of class $C^p$ everywhere on $\mathcal{M}$, we  say that $f$ is of class $C^p_q$, and we define
\begin{subequations}
    \begin{align}
        f_{\pm}(y^a)&:=\lim_{y^1\rightarrow0^{\pm}} f(y^1,y^a), \\
        \left[f\right](y^a)&:=f_+(y^a)-f_{-}(y^a),
    \end{align}
\end{subequations}
where $y^a=(y^0,y^2,y^3) $ are intrinsic  coordinates  on $\Sigma$ (in this section, lowercase Latin indices from the beginning of the alphabet, $a,b,\dots=0,2,3$, are used to denote quantities tangent to $\Sigma$).

A solution of the EC field equations \eqref{EC-equations-1}   in the region $y^1<0$ (resp. $y^1>0$) is said to match a solution in the domain $y^1>0$ (resp. $y^1<0$) if, in any appropriate coordinate system, the following hold: (i) the metric  $g_{\mu\nu}$ is of class $C^0_3$; (ii) the spin density tensor $\tensor{\tau}{_{\alpha\beta}^\lambda}$ is of class $C^{-1}_2$; (iii) the  canonical energy-momentum tensor $\mathbb{T}^{\mu\nu}$ is of class $C^{-1}_2$. These requirements are equivalent to the identities \cite{Arkuszewski:1975fz}
\begin{subequations}\label{eq: junction conditions}
    \begin{align}
        \left[\pdv{g_{ab}}{y^1}\right]&=-\frac{2}{g^{11}}\left[\tensor{K}{_{(ab)}^1}\right],
        \label{eq: junction conditions 0}
        \\
        \left[n_\lambda \tensor{\tau}{_{\alpha\beta}^\lambda}\right]&=0,
        \label{eq: junction conditions 1}
        \\
        \chi \left[n_\lambda \tensor{\mathbb{T}}{^\lambda_\mu}\right]&=-\left[\tensor{\overline{K}}{_{\mu\nu\lambda}}\right]
n_\rho \tensor{K}{^{\nu\lambda\rho}}
-\frac{1}{2}n_\mu
\left[
\tensor{\overline{K}}{_{\lambda\nu\rho}}
\tensor{\overline{K}}{^{\rho\nu\lambda}}
\right].
        \label{eq: junction conditions 2}
    \end{align}
\end{subequations}
Here,  $ \tensor{\overline{K}}{_{\mu\nu}^\lambda}$ is the projection of the contorsion tensor  onto $\Sigma$, while  $n^\mu$  the unit spacelike  normal to $\Sigma$, which  satisfies $n_\mu n^\mu=1$ and  is supposed to be of class $C^{0}_3$.   Equations \eqref{eq: junction conditions 1} and \eqref{eq: junction conditions 2} constitute the junction conditions for the matter fields $\mathbb{T}^{\mu\nu}$ and $\tensor{\tau}{_{\alpha\beta}^\lambda}$, and represent the only independent restrictions imposed on their jumps $\left[\mathbb{T}^{\mu\nu}\right]$ and  $\left[\tensor{\tau}{_{\alpha\beta}^\lambda}\right]$.

In the special case where a vacuum external solution defined, say, for $y^1>0$ has   to be matched across $\Sigma$ to an inner gravitational field  sourced by $\mathbb{T}^{\mu\nu}$ and $\tensor{\tau}{_{\alpha\beta}^\lambda}$ in $y^1<0$,  the  metric must comply with
\begin{subequations}
\begin{align}
\left.g_{\mu\nu}\right\vert_{+}&=\left.g_{\mu\nu}\right\vert_-,
\\
\left.\pdv{g_{ab}}{y^1}\right|_+
&=
\left.\pdv{g_{ab}}{y^1}\right|_-
+
\left.\frac{2}{g^{11}}K_{(ab)}{}^1\right|_- ,
    \end{align}
\end{subequations}
while the junction conditions for the matter fields reduce to
\begin{subequations}
\label{vacuum-junct-cond}
\begin{align}
\left.n_\lambda \tensor{\tau}{_{\mu\nu}^\lambda}\right\vert_-&=0,
        \label{eq: junction conditions vacuum a}
        \\
        \left.
\left(
\chi n_\lambda \tensor{\mathbb{T}}{^\lambda_\mu}
+
\frac12 n_\mu
\tensor{\overline{K}}{_{\lambda\nu\rho}}
\tensor{\overline{K}}{^{\rho\nu\lambda}}
\right)
\right|_- &=0 .
        \label{eq: junction conditions vacuum b}
    \end{align}
\end{subequations}

This  situation  arises, for example, when the interior region is filled with a Weyssenhoff fluid, with  matching conditions \eqref{vacuum-junct-cond} boiling down to
\begin{subequations}\label{eq: junction conditions W-1}
    \begin{align}
       \left. n_\mu u^\mu\right\vert_-&=0, \label{first-junct-gen}
        \\
       \left. p\right\vert_-&=\left.\chi c^2 (n_\mu \ell^\mu)^2\right\vert_-, \label{second-junct-gen}
    \end{align}
\end{subequations}
where, we recall,  $u^\mu$ is the fluid four-velocity, $p$  the pressure, and $\ell_\mu$  the Pauli-Lubanski  four-vector defined in Eq. \eqref{eq:def Pauli-Lubanski}.

\subsubsection{Application to our model} \label{Sec:junction-our-setup}

We can now apply the matching procedure to the  slowly rotating Weyssenhoff fluid studied in this paper. 

We introduce coordinates $y^\alpha=(cT,R,\Theta,\Phi)$, with $T$   timelike  and $(R,\Theta,\Phi)$  spacelike, and let the timelike  hypersurface $\Sigma$   separating the matter interior  from the exterior vacuum region be defined by $R=0$, with  $y^a=(cT,\Theta,\Phi)$ the intrinsic coordinates.  Owing to the symmetries of our model, $\Sigma$ is invariant under translations in $T$ and $\Phi$. In the  coordinates $y^\alpha$, the normal (co)vector is given by
\begin{align}
    n_\mu=(0,n_R,0,0),
\end{align}
and hence the Weyssenhoff-fluid junction conditions \eqref{eq: junction conditions W-1} take the  form
\begin{subequations}\label{eq: junction conditions W-2}
    \begin{align}
       \left. u^R\right\vert_-&=0, \label{eq: junction conditions W-2-1}
        \\
     \left.   p\right\vert_-&=\left.\chi c^2 \left(n_R \ell^R\right)^2\right\vert_-. \label{eq: junction conditions W-2-2}
    \end{align}
\end{subequations} 

The first, Eq. \eqref{eq: junction conditions W-2-1},  is automatically met in our settings. In fact, recall that in  spherical coordinates  $\tilde{x}^\mu=(ct,r, \theta,\varphi)$ the fluid four-velocity  is (cf. Eq. \eqref{four-velocity-spherical-aux})
\begin{align}
    \tilde{u}^\mu=(\tilde{u}^0,0,0,\tilde{u}^\varphi).
\end{align}
The symmetries permit the identifications  $T \equiv t$ and $\Phi \equiv \varphi$ (and  $R\equiv r-R_\star$, $\Theta\equiv\theta$ in the spherically symmetric limit; see below). Therefore,  the only nonzero Jacobian components between coordinates  $\tilde{x}^\mu=(ct,r, \theta,\varphi)$ and $y^\alpha=(ct,R,\Theta,\varphi)$ are in general
\begin{align}
    \pdv{R}{r},\,\,\pdv{R}{\theta},\,\,\pdv{\Theta}{r},\,\,\pdv{\Theta}{\theta}.
\end{align}
Therefore $\tilde{u}^r=\tilde{u}^\theta=0$ implies  $u^R=u^\Theta=0$, and then Eq. \eqref{eq: junction conditions W-2-1} is satisfied.

Relation  \eqref{eq: junction conditions W-2-2}   constrains  the \emph{a priori} arbitrary profiles   of the source pressure and spin density, and  can be fulfilled only once the internal properties of the matter distribution are specified, typically via an equation of state. More importantly, it shows that the spin density tensor, as well as the pressure, can exhibit a jump discontinuity across the Weyssenhoff fluid-vacuum interface. In these cases,  the divergence theorem cannot be applied to the integral \eqref{eq:total spin mass of the source}, and consequently  $M_s$ is, in general, nonvanishing.

To conclude, let us note that in the special scenario of the static Weyssenhoff-fluid sphere   examined in Refs. \cite{arkuszewski1974linearized,Arkuszewski:1975fz},  one may take $(T,R,\Theta,\Phi)\equiv(t,r-R_\star,\theta,\varphi)$ and perform the  matching  at the stellar surface, with  $\Sigma$ thereby defined by $r=R_\star$ (note that $\Sigma$ is not at $r=0$, which would be the center). Under these hypotheses, the first junction identity \eqref{first-junct-gen} is readily obeyed,  because the four-velocity takes the simplified form   $\tilde{u}^\mu=(\tilde{u}^0,0,0,0)$. The second relation \eqref{second-junct-gen}, instead, reduces to the nontrivial Arkuszewski-Kopczy\'nski-Ponomariev pressure condition \cite{Arkuszewski:1975fz}.

\section{Phenomenology and potential experimental tests}\label{sec:tests-GEM}

As discussed before,  in EC gravity torsion is intrinsically bound to matter and hence vanishes  in vacuum. Consequently, outside the material source the spacetime geometry is formally Riemannian and  test bodies obey the standard GR dynamics associated with the external metric. Nevertheless, the  exterior  perturbation \eqref{eq:final metric total} retains information about the internal structure of the source, as the effective mass receives a spin-orbit correction proportional to $M_s$, while the gravitomagnetic sector depends on the combination $\Vec{J}+2\Vec{S}$.

In this section, we explore some possible empirical consequences of these modifications and  discuss potential ways to test them. Since the external geometry has the form of a generalized Lense-Thirring metric, the most natural observables are those sensitive to weak-field gravitomagnetic phenomena. We therefore analyze the precession of classical test gyroscopes (Sec. \ref{sec:classical test gyros}) and the Sagnac effect (Sec. \ref{sec:Sagnac effect}). Last,  we explore possible connections with spin-polarization mechanisms such as the Barnett and Einstein-de Haas effects (Sec. \ref{sec:Barnett}).

\subsection{Precession of classical test gyroscopes}\label{sec:classical test gyros}

In the context of GR, the  motion of an isolated \textit{extended}  object endowed with a \textit{macroscopic}  angular momentum is described, within the pole-dipole scheme,  by the \textit{Mathisson-Papapetrou-Dixon equations}, supplemented by an appropriate spin supplementary condition \cite{Papapetrou:1951pa,Corinaldesi:1951pb,Schiff:1960gi,Dixon:1970zzaI,Dixon:1970zzII,Dixon:1974xozIII,Bini:2009zz,DeFelice:2010uvx,Steinhoff:2009tk,Steinhoff:2012rw}. By adopting the Chicone-Mashhoon-Punsly approximation \cite{Chicone2005,Singh2015a}, these equations can be  consistently expanded  to first order in the classical spin by assuming that the body has: (i) M\o{}ller radius much smaller than the local curvature radius; (ii)   four-momentum and four-velocity vectors nearly parallel. Under these hypotheses, the worldline of the object acquires a small spin-curvature acceleration producing a    slight deviation from the geodesic path, and the angular momentum  undergoes Fermi-Walker transport. This is precisely the framework suitable for studying  the  dynamics of an ideal gyroscope, i.e., a rigid, torque-free,   massive \textit{pointlike test object with macroscopic spin} $\vec{L}$ \cite{DeFelice:2010uvx,MTW1973}.  In this limit, the trajectory of the gyroscope can be assigned independently of the spin dynamics, which is governed by the precession law \cite{Ciufolini:book,MTW1973,DeFelice:2010uvx,Giovinetti_master_thesis}
\begin{align}
\dv{\lambda}\vec{L}=&\vec{\Omega}_p\times\vec{L},\label{precession-eq}
\end{align}
where  $\lambda$ is the proper time of an observer comoving with the body and $\vec{\Omega}_p$ the precession angular velocity evaluated with respect to distant stars.

Let us now consider the scenario in which the gyroscope is immersed in a weak gravitational field generated by a distribution of slowly moving matter. In this regime, the linearized spacetime metric, written in Lorentz gauge, takes the standard gravitoelectromagnetic form  \cite{MTW1973,Mashhoon1999,Barros2003,Mashhoon2003a,Mashhoon2008a,Ciufolini:book} 
\begin{align}
\dd s^2
=&   -\left(1+2\frac{\phi}{c^2}\right)c^2\dd t^2+\frac{4}{c} \left(\vec{A}\cdot  \dd\vec{x}\right)  \dd t
\nonumber\\
&+\left(1-2\frac{\phi}{c^2}\right)\dd\vec{x}\cdot \dd\vec{x},
\label{eq:Lense-Thirring metric}
\end{align}
where  $\phi$ and $\vec{A}$ are the gravitoelectric scalar potential and the
gravitomagnetic vector potential, respectively, and we have employed the  nearly Lorentzian coordinate system $x^\mu=(ct,\vec{x})$, with $\vec{x}=(x,y,z)$. 

In this setup,  it is possible to show that  $\vec{\Omega}_p$ consists of three pieces, namely  \cite{MTW1973,Ciufolini:book,Giovinetti_master_thesis}: 
\begin{subequations}
\begin{align}
\vec{\Omega}_p=&\,\vec{\Omega}_{\rm Th}+\vec{\Omega}_{\rm dS}+\vec{\Omega}_{\rm LT},\label{eq:precession tot}
\end{align}
known as  the \textit{Thomas}, \textit{de Sitter} (or \textit{geodetic}) and \textit{Lense-Thirring angular velocities}, which are given by 
\begin{align}
\vec{\Omega}_{\rm Th}=&-\frac{1}{2}\frac{\vec{\mathcal{V}}}{c^2}\times\vec{\mathcal{A}},\label{eq:precession Thomas}\\
\vec{\Omega}_{\rm dS}=&-\left(1+\frac{1}{2}\right)\frac{\vec{\mathcal{V}}}{c^2}\times\vec{\nabla}\phi,\label{eq:precession de Sitter}\\
\vec{\Omega}_{\rm LT}=&-\frac{1}{c}\vec{\nabla}\times{\vec{A}}. \label{eq:precession Lense-Thirring}
    \end{align}
\end{subequations}
Here,  $\vec{\mathcal{V}}\equiv \dd \vec{x}/\dd t$ and $\vec{\mathcal{A}}\approx \dd \vec{\mathcal{V}}/ \dd t+\vec{\nabla}\phi$ are the velocity and the nongravitational part of the acceleration of the gyroscope (so that for a freely falling motion $\vec{\mathcal{A}}=\vec{0}$).  For a detailed discussion on the operative meaning of $\vec{\Omega}_p$ and  the various terms it comprises  we refer the  reader to Refs. \cite{Ciufolini:book,MTW1973}.

Since the exterior metric \eqref{eq:final metric total} of the slowly rotating Weyssenhoff fluid exhibits a generalized Lense-Thirring structure, we can compare it with Eq. \eqref{eq:Lense-Thirring metric}, and in this way  we find the identifications (recall that $r:= \vert \vec{x} \vert$)
\begin{align}
\phi \equiv&-\frac{G}{r}(M-2M_s),\\
\vec{A} \equiv&\frac{G}{c}\frac{\vec{x}}{r^{3}}\times{(\vec{J}+2\vec{S})},
\end{align}
where we have taken into account Eq. \eqref{replacement-param}.
Hence 
\begin{align}
    \vec{\Omega}_{\rm dS}=&-\frac{3G}{2c^2}(M-2M_s)\frac{\vec{\mathcal{V}}\times\vec{x}}{r^3},\\
    \vec{\Omega}_{\rm LT}=&-\frac{G}{c^2r^3} \left[(\vec{J}+2\vec{S}) - 3 \frac{(\vec{J}+2\vec{S})\cdot\vec{x}}{r^2}  \, \vec{x}\right],
\end{align}
while we recall that  $\vec{\Omega}_{\rm Th}$ can be of the same order as $\vec{\Omega}_{\rm dS}$ for an Earth-bound observer, whereby  $\vec{\mathcal{A}}\approx\vec{\nabla}\phi=-\vec{g}_\oplus$.

The de Sitter and Lense-Thirring precessions are among the main observables in weak-field gravity. The space-based mission Gravity Probe B directly measured these precessional signatures using cryogenic  gyroscopes \cite{PhysRevLett.106.221101,Ciufolini:2004rq,Will:2011ke}, while the LAGEOS and LARES missions probed the Lense-Thirring effect through precision laser ranging determinations of satellite orbital precessions \cite{Ciufolini2013a,PhysRevD.91.044012,Ciufolini:2023czv}. Related phenomena can also be investigated through precise observations of the motion of celestial bodies \cite{Ciufolini:book}, for instance via lunar laser ranging techniques \cite{Murphy2012}. These essentially consist in evaluating  the time of flight of laser signals emitted from the Earth and retroreflected by  devices placed on the Moon,  and yield sub-centimeter constraints on the Earth-Moon dynamics  \cite{DellAgnello2012a,Battista2014,Battista2015}.  
In this context, experiments such as MOONLIGHT, together with data-analysis frameworks like the Planetary Ephemeris Program (\textit{PEP}) \cite{Bargiacchi2025,Muccino2025,Molli2026}, can play an important role in testing torsion-induced corrections in the weak-gravity limit.

\subsection{The Sagnac effect}\label{sec:Sagnac effect}

While in previous section  we have dealt with the phenomenology of mechanical gyroscopes, we now turn our attention  to  optical gyroscopes, i.e.,   devices based on the \textit{Sagnac effect} \cite{Post_RevModPhys.39.475,DARRIGOL2014789}. 

This phenomenon  is associated with the propagation of photons in  stationary, rotating spacetimes. In this setting,  consider an observer $\mathcal O$ emitting two light rays traveling in opposite directions along a  closed spatial loop $\mathcal{S}$. In general,  their roundtrip (proper) times   differ by an amount  $\Delta \tau_S$,   called \textit{Sagnac time delay}. This quantity involves the off-diagonal metric components $g_{0i}$ of the background geometry according to the formula \cite{Ciufolini:book,Kajari:2009qy,Bosi_PhysRevD.84.122002,Frauendiener2018,Giovinetti_master_thesis}
\begin{align}
\Delta \tau_S=&-\frac{2}{c}\sqrt{-g_{00}(p)}\int_{\mathcal{S}}\frac{g_{0i}}{g_{00}}\dd l^{i}
\nonumber\\
=&-\frac{2}{c}\sqrt{-g_{00}(p)}\int_{\varphi_0}^{\varphi_0+2\pi}\restr{\frac{g_{0i}}{g_{00}}}{\mathcal{S}}\dv{l^i}{\varphi}\dd\varphi \ , \label{eq:Sagnac time delay}
\end{align}
where $p$ denotes the position of $\mathcal O$ on the  path $\mathcal{S}$ and $\dd l^i=(\dd l^i/\dd \varphi)\dd\varphi$  the infinitesimal coordinate displacement along $\mathcal{S}$. 

If $\mathcal S$ is sufficiently small compared with the typical curvature scale of the spacetime, then the background metric can be locally expanded  around the  worldline of $\mathcal{O}$ by using  proper coordinates, and curvature corrections can be neglected.  In this limit,    Eq. \eqref{eq:Sagnac time delay}  reduces to
\begin{align}
    \Delta\tau_S=\frac{4}{c^2}\vec{\Omega}\cdot\vec{\mathscr{A}},\label{eq:Sagnac time delay-compact}
\end{align}
with $\vec{\mathscr A}$  the oriented area vector enclosed by $\mathcal S$ and $\vec\Omega$  the local rotation vector of the observer spatial frame (i.e., the rotation vector  characterizing the off-diagonal metric components in the proper rotating frame of   $\mathcal O$; cf. Eq. (1.88) in Ref. \cite{Maggiore:2007ulw}). The vector $\vec{\Omega} $ is related to the  precession angular velocity $\vec{\Omega}_p$ introduced in  Eq. \eqref{precession-eq}  through the identity
\begin{align}
    \vec{\Omega}=\vec{\Omega}_{\infty}-\vec{\Omega}_p,\label{eq:local rotation in terms of precession}
\end{align}
where $\vec{\Omega}_{\infty}$ is the angular velocity of $\mathcal{O}$ with respect to the fixed stars, as measured by an observer at infinity (in practice, $\vec{\Omega}_{\infty}$ can be determined via  very-long-baseline interferometry techniques \cite{Ciufolini:book}).

Although   formulas  \eqref{eq:Sagnac time delay} and \eqref{eq:Sagnac time delay-compact} are derived in the context of GR, they  remain valid  also in EC gravity, where  no (minimal) coupling between the Maxwell field and the torsion is allowed \cite{Hehl:1976kj}. Therefore, Maxwell equations retain the same form as in GR and the geometric optics limit   can be carried out  in an analogous manner  \cite{MTW1973,PRASANNA197517}. As a consequence,  light rays travel on null geodesics of the spacetime metric also in EC gravity, and the computation of Eqs. \eqref{eq:Sagnac time delay} and \eqref{eq:Sagnac time delay-compact} proceeds identically; further details can be found in Ref. \cite{PhysRevD.95.061501}.  Therefore,   for the  rotating Weyssenhoff source developed in this paper, $\Delta \tau_S$ can be computed directly by plugging  the generalized Lense-Thirring metric \eqref{eq:final metric total},  or equivalently  the  replacements \eqref{replacement-param}, into the formulas \eqref{eq:Sagnac time delay}-\eqref{eq:local rotation in terms of precession}.

Sagnac effect thus configures as another means to probe torsion effects in weak gravitational fields. The devices that measure rotations through such phenomenon are known as \textit{Sagnac gyroscopes} \cite{Post_RevModPhys.39.475,DARRIGOL2014789,Moseley2019,DiVirgilio2021}. In particular, in a \textit{ring laser gyroscope}  \cite{Chow,Stedman:1997wm,WILKINSON19871,SchreiberReview}, which is  the type of device employed in the GINGER experiment \cite{Bosi:2020dgg,Capozziello:2021goa,Altucci:2023wdo,Bosi_PhysRevD.84.122002,Giovinetti_Frontiers_10.3389/frqst.2024.1363409,10.1117/12.3027943,DiSomma_astronomy3010003,Porzio_Femto_PhysRevLett.133.013601,Giovinetti_master_thesis}, the spatial path $\mathcal{S}$ is realized by a set of corner mirrors, while  the Sagnac time delay manifests itself in the frequency difference of counter-propagating Gaussian laser beams.

Other experiments based on different technologies, such as the fiber optic gyroscopes, and/or other interferometer configurations, like the Mach-Zehnder interferometers, evaluate the Sagnac effect either via  phase shifts of continuous laser beams or via  differences in the arrival times of light pulses or photons. Some applications even rely on quantum interference, where the Sagnac time delay modifies the coincidence detection probability and the interference visibility at the output of the interferometer \cite{Mieling_2020,Mieling:2023plz,Mieling:2024jvk,Silvestri_doi:10.1126/sciadv.ado0215,Silvestri:23}.

\subsection{The Barnett and Einstein-de Haas effects}\label{sec:Barnett}

One of the  assumptions underlying our description of the rotating Weyssenhoff fluid is that the macroscopic and microscopic angular momenta are aligned along the same axis. From a dynamical point of view, this hypothesis can be motivated by known  physical mechanisms that couple rotation and magnetization (see, for example, Refs. \cite{Dolginov1976,pfenniger2003cold}). In particular, the \textit{Barnett} \cite{Barnett_PhysRev.6.239} and the \textit{Einstein-de Haas} effects \cite{PhysRevSeriesI.26.248,EinsteindeHaas1915,Frenkel:1979} describe the exchange between macroscopic rotational motion and microscopic spin degrees of freedom, and  play a key  role in modern experimental devices (see for example, Refs. \cite{PhysRevA.111.013713,Wachter:2025sbn,hdh6-r1gy} and references therein). Both scenarios can be understood as a consequence of the conservation of the total (macroscopic plus intrinsic) angular momentum. The Barnett effect consists in the magnetization induced by the rotation of an initially unmagnetized body, whereas the Einstein-de Haas phenomenon corresponds to the onset of mechanical rotation induced by a change in magnetization. In both cases, the magnetization vector tends to align with the rotation axis.
A similar phenomenon, known as the \textit{London moment}, manifests in superconductors, where it constitutes the physical mechanism behind the gyroscopes employed in Gravity-Probe B \cite{Muhlfelder2015}. 

The abovementioned effects not only provide a physical motivation for our assumption of aligned macroscopic and microscopic spins, but may open the possibility for  new laboratory setups aimed at testing gravitomagnetic effects in EC gravity (for general Earth-based laboratory applications see, e.g., Sec. 6.9 of Ref. \cite{Ciufolini:book}). A detailed analysis of such possibilities, however, lies beyond the scope of the present work.

\section{Discussion and conclusions}
\label{Sec:Conclusion}

In this paper, we have studied the exterior gravitational field of a source possessing both classical and quantum spins within linearized EC gravity. Specifically, we have examined  an isolated, axisymmetric Weyssenhoff fluid characterized by a macroscopic, slow rotation around the $z$-axis, with the intrinsic spin polarized along the same direction.

The geometry has been obtained  by first deriving  the linearized  stress-energy tensor $T_{\mu \nu}$ of the source, which depends  on the four time-independent functions $\rho$, $p$, $\omega$, and $s^z$, each constructed in accordance with the symmetries of the system (cf. Eqs. \eqref{eq:stress energy tensor cartesian}-\eqref{eq:stress-energy-explicit}). Then, the corresponding metric has been computed by resorting to a multipole expansion (see Eqs. \eqref{eq:final metric total} and \eqref{eq:total parameters}). At the leading nontrivial order, it configures as a generalized Lense-Thirring solution, where the off-diagonal components involve a linear combination of the orbital and intrinsic angular momenta, while the  gravitoelectric sector acquires the  additional effective mass correction $M_s$ linear in both the source spin and  angular velocity.  The latter contribution can be traced back to the spin-orbit energy density \eqref{eq:effective spin mass density}, determined by the electric piece $\vec{q}$ of the spin tensor, which  is globally nonvanishing in our setup.   

These spin-rotation couplings, which remarkably contribute to the gravitational field even in the weak-field  regime, can be  better understood if the rotating Weyssenhoff source is interpreted as the gravitational analogue of a rotating magnet. In such a system, the electric field vanishes in the magnet rest frame but appears in a rotating frame. This situation mirrors the local behavior of $\vec{q}$ and is consistent with the constraints imposed by the Frenkel condition.

The phenomenology associated with the gravitational field of the  rotating Weyssenhoff fluid has also been discussed. The new corrections can potentially affect standard weak-field observables, such as gyroscope precession and the Sagnac effect, and may be probed through both space-based and laboratory tests. In this regard,  recent developments in condensed-matter physics may provide additional tools for exploring the connection between rotation, quantum interactions, and gravity. In particular, novel routes may emerge from spin-polarization mechanisms such as the Barnett and Einstein-de Haas phenomena. More broadly, the growing empirical control over rotating quantum systems and quantum materials, together with the observation of rotation-induced effects in interferometric and entanglement-based setups \cite{Toros:2019ptw,Toros:2022tpq,Cromb:2022ptr,Cromb:2023zmr,Cromb:2023ozo} and studies based on fast-rotating matter  \cite{Filgueiras:2025eci,2023JPCM...35d5401B,RevModPhys.81.647}, may open new avenues for testing  spin-rotation couplings beyond traditional astrophysical and gravitational experiments.

Although our model can furnish only a preliminary account of  fast-rotating astrophysical objects,  it has  nevertheless shed light on the structure of spin-rotation couplings. At the leading order,
the quantum spin enters the metric with a characteristic factor of two relative to the orbital contribution, in agreement with the post-Newtonian analysis of compact binaries dynamics performed in Refs. \cite{Battista:2022sci,Battista:2023znv,DeFalco:2023djo}. Developing the present framework further through approaches like the  post-Newtonian or the post-Minkowskian approximation schemes might reveal additional intriguing facets of the  interplay between macroscopic and microscopic spins, potentially leading to new unexpected results. This is a worthwhile direction for future research.

\section*{Acknowledgements}
The authors acknowledge the support of Istituto Nazionale di Fisica Nucleare (INFN), {\it Iniziative Specifiche} MOONLIGHT-2,  GINGER, and QGSKY.

\bibliography{references}

\end{document}